\documentclass[aps,pra,10pt,twocolumn,superscriptaddress,nofootinbib,notitlepage,showpacs,floatfix,nobalancelastpage,longbibliography]{revtex4-2}

\usepackage[english]{babel}

\usepackage{graphicx,graphics,epsfig,subfigure,times,bm,bbm,amssymb,amsmath,amsfonts,mathrsfs}
\usepackage[scr=boondoxo,scrscaled=1.05]{mathalfa}
\usepackage[matrix,frame,arrow]{xypic}
\usepackage[pdftex]{color}
\usepackage{extarrows}
\usepackage[pdfstartview=FitH,colorlinks=true, allcolors=blue]{hyperref}
\usepackage{cancel}

\newtheorem{theorem}{Theorem}

\long\def\ignore#1{}

\newcommand{\beq}{\begin{equation}}
\newcommand{\eneq}{\end{equation}}
\newcommand{\beqnn}{\begin{equation*}}
\newcommand{\eneqnn}{\end{equation*}}
\newcommand{\beqy}{\begin{eqnarray}}
\newcommand{\eneqy}{\end{eqnarray}}
\newcommand{\beqynn}{\begin{eqnarray*}}
\newcommand{\eneqynn}{\end{eqnarray*}}

\newcommand{\braket}[1]{ \langle  #1 \rangle  }
\newcommand{\ket}[1]{ | #1 \rangle  }
\newcommand{\bra}[1]{ \langle  #1 | }

\newcommand{\Tr}{\mathrm{Tr}}
\newcommand{\subfigimg}[3][,]{%
  \setbox1=\hbox{\includegraphics[#1]{#3}}
  \leavevmode\rlap{\usebox1}
  \rlap{\hspace*{2pt}\raisebox{\dimexpr\ht1+0.2\baselineskip}{#2}}
  \phantom{\usebox1}
}
\newenvironment{proof}[1][Proof]{\noindent\textbf{#1.} }{\ \rule{0.5em}{0.5em}}
\addto\captionsenglish{}

\begin{document}

\title{
Broadband spectroscopy of quantum noise 
}
\author{Yuanlong Wang}
\affiliation{Centre for Quantum Computation and Communication Technology (Australian Research Council), Centre for Quantum Dynamics, Griffith University, Brisbane, Queensland 4111, Australia}
\affiliation{Key Laboratory of Systems and Control, Academy of Mathematics and Systems Science, Chinese Academy of Sciences, Beijing 100190, People's Republic of China}

\author{Gerardo A. Paz-Silva} \thanks{g.pazsilva@griffith.edu.au}
\affiliation{Centre for Quantum Computation and Communication Technology (Australian Research Council), Centre for Quantum Dynamics, Griffith University, Brisbane, Queensland 4111, Australia}

\begin{abstract}
Characterizing noise is key to the optimal control of the quantum system it affects. Using a single-qubit probe and appropriate sequences of $\pi$ and non-$\pi$ pulses, we show how one can characterize the noise a quantum bath generates across a wide range of frequencies -- including frequencies below the limit set by the probe's  $\mathbb{T}_2$ time. To do so we leverage an exact expression for the dynamics of the probe in the presence of non-$\pi$ pulses, and a general inequality between the symmetric (classical) and anti-symmetric (quantum) components of the noise spectrum generated by a Gaussian bath. Simulation demonstrates the effectiveness of our method.

\end{abstract}

\maketitle

\section{Introduction}

Characterizing noise is key not only to the implementation of high fidelity operations in quantum devices, but to enable detailed understanding of the physical processes inducing such noise \cite{dowling2003,preskill2018}. Protocols capable of providing such information infer it from the measured response of a quantum system to probing control (which need not be unitary) in the presence of the noise one desires to characterize. Depending on the purpose and control capabilities, different protocols can be used, mostly under the term \textit{quantum noise spectroscopy} (QNS). For example, relatively costly frequency comb-based \cite{alvarez2011} or Slepian control \cite{norris2018} protocols can in principle sample frequency-domain representations of the noise correlations in high detail, while \cite{vezvaee2022} reconstructs a large frequency regime of the spectrum. Efficient frame-based protocols \cite{chalermpusitarak2021,dong2023} extract only the relevant information necessary to control the system given fixed control constraints/capabilities. Others can provide information not on noise correlations, but rather a description of the process tensor describing the open system dynamics \cite{white2020}. 

Generally speaking, such protocols are perturbative in nature, in the sense that they rely on the perturbative expansion of the time-dependent behaviour of expectation values, typically as a function of the leading-order bath correlation functions\footnote{Process tensor methods are in general not perturbative, but their complexity grows exponentially with the number of interventions -- roughly speaking the ``complexity of the control''. Tractable process tensor approaches rely on perturbative-like approximations based on other criteria which we will not consider here.}. This results in either increasingly more complex and costly implementations, or less accurate/reliable reconstructions. Indeed, under certain conditions, protocols that provide a qualitative -- rather than quantitative -- description of classical and quantum noise have been recently proposed \cite{quintana2017,wang2021,shen2023}. That is, they are essentially applicable to systems where the noise-inducing environment is weakly coupled to the system of interest. While this can be a reasonable assumption for systems where the weak coupling is a necessity and has been in a sense hard-coded in its physical description and design, e.g. for quantum computing purposes, it may not be for a general system where a strong system-bath coupling regime is of interest, e.g., for sensing if the environment is the object of interest or in the road to designing a low noise physical system.  

A notable exception to this is when the physical model of interest admits an exact solution in the presence of the probing control. In such cases, no weak coupling assumption is necessary, allowing one, in principle, to study the problem in typically inaccessible regimes, e.g., in the long-time regime. This, in turn, implies the ability to study the noise in great detail. Particularly relevant to this paper, this allows one to analyze the effect of very low frequency noise \cite{paladino2014}, which is inherently difficult for methods relying on perturbative expansions and thus on ``total evolution time"-bounded experiments. This is of interest beyond the full characterization scenario, for example opening the possibility to wide-range bath thermometry, which would not be possible otherwise. 

In practice, however, few exactly solvable models are known. Prominent among these is the case of dephasing Gaussian noise in the presence of instantaneous $\pi$-pulse control \cite{norris2016,pazsilva2017,norris2022} -- described in detail in Eq.~\eqref{basicH}. The drawback of this model is that some components of the noise do not contribute to the $\pi$-pulse-modulated dynamics and are thus invisible to the system qubit. Indeed, one can show that only the symmetric component of the bath correlations, i.e., the classical component (dubbed $c$ in this paper), contribute, while the antisymmetric, i.e., the quantum component (dubbed $q$), disappear from the equations. In other words, in the above scenario the quantum nature of the bath is invisible. And yet, it not only would contribute when more general control is applied, but also contains highly coveted information about the physical source of such noise. It is worth highlighting in view of recent results that the above is true for the common type of coupling we will use here -- see Ref. \cite{pazsilva2017} for more exotic coupling types whose dynamics reveals the presence of the antisymmetric component of the bath correlations -- and under the assumption that the bath state at initialization (time zero) is independent of the system state, i.e., an initialization procedure can prepare $\rho_S \otimes \rho_B$ for a set of $\rho_S$ but fixed $\rho_B.$ When the latter is not the case~\cite{wang2021}, the resulting effective model can be mapped to one of the exotic couplings mentioned, and thus depends on the quantum component of the noise.

In this paper, we show how one can leverage non-$\pi$ pulses to extract information about the quantum component of the noise while using non-perturbative equations for the dynamics. In addition to the existing capabilities granted by $\pi$-pulse-only control, the new paradigm allows us to study both classical and quantum components in the elusive strong coupling, long evolution time regime. Further, we propose a systematic method to selectively reconstruct any frequency range of interest of the noise spectra based on the observed single-qubit reduced dynamics. Moreover, we establish a novel inequality bounding the strength of the quantum spectrum by the classical counterpart, which for the first time clarifies the relationship between the two spectra.

The structure of this paper is as follows. Sec. \ref{sec2} summarizes the basic model, motivation and results of this paper. Sec. \ref{sec:basic} derives and analyzes the exact solution for the reduced qubit dynamics in the time domain. Sec. \ref{sec:freq} moves to the frequency domain and presents a systematic method reconstructing the $c$ and $q$ spectra, together with a novel inequality characterizing their quantitative relationship. Sec. \ref{sec:simu} provides simulation results to validate our spectra reconstruction algorithm. Conclusions are presented in Sec. \ref{sec:con}.

\section{Key elements and results}\label{sec2}

Let us start by summarizing some of the key elements and results in this paper, while leaving the derivations to the later sections. 

We consider a controlled single qubit probe $S$ dephasingly coupled to an uncontrollable bath $B$. In the interaction picture associated with the bath Hamiltonian $H_B$, the joint system-bath dynamics is governed by a Hamiltonian of the form
\begin{equation}
\label{basicH}
  H(t)=\sigma_z\otimes B(t)+H_{\rm{ctrl}}(t)\otimes I_B,  
\end{equation}
where $B(t)$ is a bath operator and $H_{\rm{ctrl}}(t)$ is the control Hamiltonian.

The $B(t)$ and the initial state of the bath $\rho_B$ are assumed to lead to noise that is zero-mean, Gaussian and stationary. That is, noise which is entirely described by the second-order cumulant, i.e., $C^{(2)} (B(t),B(t')) = \langle{\mathrm{Tr}}_B[{B(t) B(t') \rho_B}]\rangle_c = \langle{\mathrm{Tr}}_B[{B(t-t') B(0) \rho_B}]\rangle_c \equiv \langle B(t-t') B(0) \rangle$, where the expectation $\braket{\cdot}$ includes both the classical ensemble average $\braket{\cdot}_c$ and the quantum expectation ${\mathrm{Tr}}_B({\cdot \rho_B})$. A paradigmatic example of this noise is the one generated by a bosonic bath, where $B(t)$ is linear in the creation and annihilation operators and $\rho_B$ is a thermal state with inverse temperature $\beta$. 

We will further assume that the control Hamiltonian is capable of implementing instantaneous $\theta$ rotations around a given axis -- here $\sigma_y$ for concreteness. We will assume a minimum time of $\Delta$ between pulses and control time resolution $\delta$, i.e., if the position of the $i$-th pulse is denoted by $t_i$, then $t_i = k \delta$ and $t_{i+1}- t_i= K \delta$ for integers $k,K$. Since $K\delta\geq \Delta$, we must have $K\geq \lceil \Delta/\delta\rceil$. Since the action of $\pi$-pulses preserves the dephasing character of the Hamiltonian, it will be useful to divide the control Hamiltonian into a part that solely executes $\pi$-pulses and the complement, i.e. $H_{\rm ctrl}(t)= H_{\rm ctrl}^\pi(t) + H_{\rm ctrl}^{\neg \pi}(t).$ 
In this way, introducing the propagator $U_{\rm{ctrl}}^\pi(t)\equiv \mathcal{T}_+\exp(-{\rm i} \int_0^t ds H_{\rm{ctrl}}^\pi(s))$, the interaction picture Hamiltonian with respect to $H_{\rm{ctrl}}^\pi(t)$ (i.e., the ``toggling frame") yields
\begin{align}
    \label{Hinte}
    H_I(t)&=U_{\rm{ctrl}}^\pi(t)^\dagger [H(t)-H_{\rm{ctrl}}^\pi(t)]U_{\rm{ctrl}}^\pi(t)\notag \\
    &=y(t)\sigma_z\otimes B(t)+H_{\rm{ctrl}}^{\neg \pi}(t)\otimes I_B,
\end{align}
where $y(t)$ is the so-called ``switching function" taking values $\pm 1$. The non-$\pi$ pulses play a dual purpose in our protocol. First, as is common in Ramsey-type experiments, as a way to prepare the initial state. Second, and more importantly for this paper, to break the dephasing symmetry in the toggling-frame Hamiltonian. 

Under these conditions, one can show (see Sec. \ref{sec:basic} for details) that the exact dynamics of the qubit at time $T$ depends on the integrals
$$
\label{basicInte}
I^\pm (T) = \int_{0}^{T} dt \int_{0}^{t} dt' Y^-(t) Y^\mp (t') C^\pm(t,t') ,
$$
where we have used the short-hand notation $C^\pm(t,t') \equiv \langle [B(t),B(t')]_\pm \rangle$, with $[A,B]_\pm\equiv AB\pm BA$, to denote the classical (+) and quantum (-) components of the cumulants describing the noise statistics. We will refer to their Fourier transform $S^\pm(\omega)$ as the $c$ and $q$ spectra. Additionally, here the $Y^\pm(t)$ are user-defined switching functions taking values in $\{-1,0,1\}$ that satisfy what we dub the {\it incompatibility condition}, namely
\begin{equation}\label{incop} |Y^+(t)+ Y^-(t) |= 1 \textrm{    and     }   Y^+(t) Y^-(t) =0.
\end{equation}

\subsection{Consequences of dephasing-preserving control}
Since a key ingredient of our scheme is breaking the dephasing preserving symmetry, it is worth understanding some of the constraints that arise due to such strong symmetry:

(i) {\it The qubit is insensitive to the $q$ component of the noise.-} In the dephasing-preserving scenario one finds that $Y^-(t) = y(t) \in \{-1,1\}$ and $Y^+(t) = 0,$ so that $I^- (T) = 0$ and 
\begin{align*}
I^+ (T) &= \frac{1}{2} \int_{0}^{T} dt \int_{0}^{T} dt' y(t) y(t')  C^+(t,t'),
\end{align*}
i.e., only the classical component of the bath contributes to the dynamics. This can be overcome in at least two ways: using multiple qubits or using non dephasing-preserving control. The former implies a considerable increase in physical resources and control complexity, e.g., entangled probes or entangling operations, but has the distinct advantage of admitting exact equations~\cite{pazsilva2017}, while the latter can no longer be studied via exact equations and thus is limited to short-time/weak-coupling~\cite{pazsilva2019} or steady state~{\cite{quintana2017}} regimes.

This limitation in turn implies the inability to extract the information encoded in the $q$ spectra or in the relationship between the $c$ and $q$ spectra, such as the temperature of the thermal bath in the bosonic bath case.

(ii) {\it The noise frequency which can be sampled is effectively lower bounded.-} Even if one is only interested in the $c$ spectrum, the dephasing-preserving scenario leads to sampling limitations. To see this, let us expand the $c$ (stationary) correlation function in Fourier series so that
\begin{equation}\label{lowfeq}
C^+(\tau = t-t',0) = \sum_{k=0}^\infty c_k \cos( k \omega_0 \tau),
\end{equation} 
for some small frequency resolution $ 0< \omega_0 \ll 1$. Extracting the ``low frequency'' information implies the need to learn  $\{c_k\}$ for $k \in \{0,1,\cdots, K_{\rm min}\}$. Even more, assume the values of $c_k$ for $k > K_{\min}$ and the function value of $C^+(\tau ,0)$ from $\tau=0$ up to some $T_{\rm{s}}$ are known. The problem is then to generate a system of (linear) equations out of Eq.~\eqref{lowfeq} from which the unknown $\{c_k\}_{k\leq K_{\rm min}}$ can be reliably estimated.   

Analysing the relative difference between  the ``coefficients'' associated to $c_k$ and $c_{k+1},$ i.e., the corresponding cosines, one finds that for small $k \omega_0 \tau,$
\begin{equation*}
\begin{aligned}
\cos(k \omega_0 \tau ) - \cos[ (k+1) \omega_0 \tau ] &=2\sin[(k+\frac{1}{2})\omega_0 \tau]\sin\frac{\omega_0 \tau}{2}\\
&\sim (k+\frac{1}{2})2\omega_0 \tau\sin\frac{\omega_0 \tau}{2}.
\end{aligned}
\end{equation*}
This shows that differentiating between two subsequent $c_k$s becomes harder the smaller $k$ and $\omega_0$ are, unless $\tau$ can be made large, say $\tau > T_{\rm upper}.$
 
On the other hand, the contribution of $C^+(\tau>T_{\rm upper},0) $ to the dynamics is contained in an integral like $ \bar{I}^+|_{T_{\rm upper}} \equiv \int_{T_{\rm upper}}^{T} dt \int_{0}^{\eta} dt' Y^-(t) Y^- (t') C^+(t-t',0)$, for some small $\eta>0$, mixed with the full history of the probe $I^+ (T)$. Indeed, measuring the expectation value of $\sigma_x$ at a time $T$ given an initial  state $\ket{+}_x$ (the eigenstate of $\sigma_x$ corresponding to $+1$) leads to 
\begin{align*}
E[ \sigma_x (T)] _{\ket{+}_x}  & = e^{ - I^+ (T) } \\
& = e^{- \bar{I}^+|_{T_{\rm upper}}} \times e^{-I^+ (T)|_{\rm rest}},
\end{align*}
i.e., the signal associated to $C^+(\tau > T_{\rm upper},0)$ is obscured by the decay from the history of the dynamics $e^{-I^+ (T)|_{\rm rest}}$ and its contribution is thus harder to discriminate experimentally the larger $T_{\rm upper}$ -- the smaller the target frequency range -- is. 

Additionally, the inability to accurately and broadbandly sample also implies the inability to extract reliable information about physical parameters which lie within certain ranges. For example, characterizing low bath temperatures (high $\beta$ regime) in the bosonic example we alluded to earlier becomes impractical.

\subsection{Beyond dephasing-preserving control}

Overcoming these limitations will be the main contribution of this work, and we will do so by exploiting the ability to generate effective switching functions with zero values in set domains. 

One way to do so, is to use incoherent mechanisms. This was done for the purely classical bath scenario in \cite{sakuldee2020,sakuldee2020b,fink2013}, by interleaving $n$ blocks of preparation, noisy evolution and measurement, with {\it idle} periods. Importantly, in this idle periods the qubit {\it effectively} does not interact with the bath or rather the effect of the noise on the qubit during those idle periods is deleted in the preparation of the following block. In this way, these protocols eventually allow one to sample the low frequency region of the (classical) noise spectrum. 

In the quantum bath scenario, however, the situation is different. While a preparation would indeed delete the effect of the bath on the probe qubit, the opposite is not true. The presence of the qubit influences the bath, and the final expectation values will reflect this. 

To overcome this limitation we propose to use instead coherent control and a detailed analysis of the system evolution in the presence of the quantum bath. That is, we exploit $H_{\rm ctrl}^{\neg \pi}(t),$ instead of sequential measurements. WE observe that when control is not dephasing-preserving $Y^+(t)$ need not vanish, and $Y^\pm(t)$ take values in $\{-1,0,1\}.$ This allows one to build combinations of expectation values of operators which are only sensitive to part of the history, in the sense that they only depend on integrals of the form
$$\mathcal{I}^\pm (\vec{T})  =  \int_{{T_3}}^{{T_4}} dt \int_{{T_1}}^{{T_2}} dt' y(t) y'(t') C^\pm(t-t',0),
$$
for useful (but not quite arbitrary) values of $0 \leq T_i \leq  T$ and $y(t),y'(t) \in \{-1,1\}.$ In other words, even in the quantum bath scenario, our protocol allows combinations of expectation values which yield quantities that mimic the 'idle' behaviour in the protocols mentioned earlier. Section \ref{sec:basic} focuses on deriving this result and on showing how these integrals can be isolated by appropriate combinations of observables, control sequences, and initial states. Importantly, we shall show how to do so exactly, i.e., without invoking any approximation, allowing us to characterize the effect of both $q$ and $c$ baths in a quantitatively accurate way, in contrast with recent results \cite{shen2023}.

With this ability in hand, we achieve {\it broadband} characterization of the noise (both $c$ and $q$ components), i.e., even in the long-time  \&  strong-coupling regime.  In particular, we demonstrate how this ability allows us to characterize the  {\it very low frequency range} of both the $c$ and $q$ spectra of a general stationary Gaussian bath -- below the one set by the $\mathbb{T}_2$  time of the probe.

\section{Time-domain derivation}\label{sec:basic}
\subsection{Reduced qubit dynamics}\label{subsec:dynamics}

While moving away from the dephasing-preserving scenario is desirable from the point of view of allowing the qubit to be sensitive to more features of the bath, a considerable complication is added to the analysis: the ability to write exact analytical expressions for the response of the qubit to control and the noise is lost (or at the very least severely compromised). This leads to the need for perturbative approaches, which in turn restrict the ability to extract information which is encoded in the long-time behaviour of the qubit, e.g., low frequency features of the noise. In terms of going beyond $\pi$-pulses, for example, Ref. \cite{laraoui2013} proposed a protocol non-trivially employing non-$\pi$ pulses to reconstruct a $c$ spectrum with a simplified structure, which leverages many approximations in the deduction, but does not develop a systematic general solution. To overcome these sort of complications, we derive an exact solution for the reduced qubit dynamics subject to general non-$\pi$ pulses, laying a foundation for resolving both the $c$ and $q$ spectra. While the scaling of the complexity of the solution -- which may be of independent interest\footnote{The expression derived here can be related to the process tensor approach~\cite{ProcessTensor0} . This is an avenue that we will explore in a latter work.} -- is exponential in the number of non-$\pi$ pulses being considered and is thus impractical in many scenarios of interest, only a few non-$\pi$ pulses are necessary to achieve the results in this work. 

Let us start by denoting $U(t_a,t_b)\equiv\mathcal{T}_+ \exp\big[-{\rm i} \int_{t_a}^{t_b}y(t)\sigma_z\otimes B(t)dt\big]$. Let the non-$\pi$ pulse control $H_{\rm{ctrl}}^{\neg \pi}(t)=\sum_{j=0}^N\theta_j\sigma_y\delta(t-t_j)$ implement $\theta_j$ angle pulse at $t_j$, where $t_0=0$ and $t_N=T$ are the preparation and measurement times, respectively. They generate unitaries $[\theta_j]_y \equiv e^{-{\rm i} \frac{\theta_j}{2} \sigma_y}$ which can be generically written as $[\theta_j]_y=\sum_{r=0}^1 c^{(j)}_r ({\rm i} \sigma_y)^r,$ with the constraint  $\sum_{r=0}^1 (c^{(j)}_r)^2=1$. Concretely, the coefficients are related to the rotation angle via $c^{(j)}_r\equiv \cos(\frac{\theta_j}{2}+r\frac{\pi}{2}),$ for the unitary control we consider here\footnote{Notice that the decomposition can be made also if the operations/interventions are not unitary, e.g., measurements or projectors, albeit with a different constraints for the $c^{(j)}_r$.}. Dividing the full-time propagator $U(T)\equiv \mathcal{T}_+ \exp\big[-{\rm i} \int_{0}^{T}H_I(t)dt\big]$ into time intervals defined by the non-$\pi$ pulse timings, one obtains
\begin{align}
&\quad U(T)=  [\theta_{N}]_y U(t_{N-1}, t_N ) [\theta_{N-1}]_y  \cdots [\theta_1]_y U(t_0,t_1) [\theta_0]_y\notag \\
&\!=\! \sum_{\vec{r}} (\!\prod_{j=0}^{N} c_{r_j}^{(j)}\!) {\rm i}^{|\vec{r}|}    \sigma_y^{r_N} U(t_{N-1}, t_N )\sigma_y^{r_{N-1}} \cdots \sigma_y^{r_1}U(t_0, t_1)  \sigma_y^{r_0}\notag \\
 &\! =\! \sum_{\vec{r}} (\prod_{j=0}^{N} c_{r_j}^{(j)}) {\rm i}^{|\vec{r}|_0}  U_{\vec{r}} (T)  \sigma_y^{|\vec{r}|_0},
 \label{eqd1}
\end{align}
with $\vec{r}\equiv(r_0,\cdots, r_N)$ and $|\vec{r}|_k\equiv\sum_{j=k}^N r_j$. The summation $\sum_{\vec{r}}$ in Eq. \eqref{eqd1} ranges over all the combinations of $r_j\in\{0,1\}$, and for the last line we define 
\begin{align}
U_{\vec{r}}(T)&\equiv  \bar{U}(t_{N-1}, t_N )\bar{U}(t_{N-2} , t_{N-1})\cdots \bar{U}(t_0,t_1),\notag \\
\bar{U}(t_{k-1} , t_k )&\equiv\mathcal{T}_+ \exp \Big[ -{\rm i}\int_{t_{k-1}}^{t_k} dt (-1)^{|\vec{r}|_k}y(t)\sigma_z \otimes B(t) \Big].\notag 
\end{align}
In our protocols, it will be sufficient to work with up to four non-$\pi$ pulses ($N=3$ in the above equations).

Any information we extract from the system comes in the form of system measurements. The expectation value of a system-only operator $\hat O$ is given by  $E \big[\hat{O}(T)\big]_{\rho_{S}\otimes\rho_B}= \Tr \big[ U(T) \rho_S\otimes \rho_B U(T)^\dagger \hat{O} \big]=\Tr \big[ V_{\hat{O}}(T) \rho_S \hat{O} \big]$, where $V_{\hat{O}}(T)\equiv\langle  \hat{O}^{-1} U(T)^\dagger \hat{O} U(T)  \rangle$. Furthermore, the {\it system operator} $V_{\hat O}(T)$ can be calculated, by considering $\hat O\equiv \sigma_{\hat o}$ with $\hat o\in\{x,y,z\}$ and introducing the shorthand notation $f_{\hat o}^\alpha\equiv {\rm Tr}(\sigma_{\hat o}\sigma_\alpha\sigma_{\hat o}\sigma_\alpha)/2 \in \{-1,+1\}$ where $\alpha\in\{x,y,z\}$, as 
\begin{align}
  &\quad V_{\hat O}(T)  \notag \\
  &=\left\langle  \hat{O}^{-1} \sum_{{\vec{r}'}} (\prod_{j=0}^{N} c_{r'_{j}}^{(j)}) {\rm i}^{-|{\vec{r}'}|_0}  \sigma_y^{|{\vec{r}'}|_0} U_{{\vec{r}'}} (T)^\dagger   \hat{O}\right.\notag \\
  &\quad \cdot\left.\sum_{\vec{r}} (\prod_{j=0}^{N} c_{r_j}^{(j)}) {\rm i}^{|\vec{r}|_0}  U_{\vec{r}} (T)  \sigma_y^{|\vec{r}|_0}  \right\rangle \label{VO} \\
   &\equiv \sum_{\vec{r},{\vec{r}'}} {\rm i}^{|\vec{r}|_0-|{\vec{r}'}|_0} \prod_{j=0}^{N}c_{r'_{j}}^{(j)}c_{r_{j}}^{(j)} \big({f_{\hat o}^y}\big)^{|{\vec{r}'}|_0}\sigma_y^{|{\vec{r}'}|_0}V_{\hat o,\vec{r},{\vec{r}'}}(T)\sigma_y^{|\vec{r}|_0},\notag
\end{align}
from which we see that the quantity of interest 
$V_{\hat o,\vec{r},{\vec{r}'}}(T) = \langle{\hat O}^{-1} U_{{\vec{r}'}}(T)^\dagger \hat O U_{\vec{r}}(T)\rangle$ contains the information about how the noise interacts which the unitaries associated to each $\vec{r}, \vec{r}'.$

The crucial observation is that each of the $V_{\hat o,\vec{r},{\vec{r}'}}(T)$ can be written in terms of an effective {\it dephasing} Hamiltonian. Thus, given the Gaussian bath scenario we are considering, one can write an exact expression for each of them using the cumulant expansion technique \cite{kubo1962,pazsilva2016,pazsilva2017, pazsilva2019}. In more detail, $V_{\hat o,\vec{r},{\vec{r}'}}(T) $ can be written in a time-ordered exponential form
\begin{equation}\label{eqc4}
\begin{aligned}
     V_{\hat o,\vec{r},{\vec{r}'}}(T)
    =&\mathcal{T}_+\exp\Big[-{\rm i}\int_{-T}^Tdt H_{\hat o,\vec{r},{\vec{r}'}}(t)\Big]\\
    =& \exp\big[-{\rm i}\mathcal{C}^{(1)}_{\hat o,\vec{r},\vec{r}'}(T)-\frac{1}{2}\mathcal{C}^{(2)}_{\hat o,\vec{r},\vec{r}'}(T)\big].
\end{aligned}
\end{equation}
In the first equality, we have used the effective Hamiltonian 
\begin{align*}
&\quad {H}_{\hat o,\vec{r},{\vec{r}'}}(t)  \\
&=\begin{cases}
 -y_{{\vec{r}'}}(T-t)\hat O^{-1}  \sigma_z  \hat O \otimes B(T-t)  & {\rm for}\ 0< t \le T \vspace{1mm}, \\ 
\,\,\,\,\,\,\,\,\,\,\,\,\,\,\,\,\, y_{\vec{r}}(T+t) \sigma_z \otimes B(T+t)   & {\rm for}\ -T\le t \le 0,
  \end{cases}
\end{align*}
with $y_{\vec{r}}(t)$ the piece-wise function
\begin{align}\label{eqc05}
	y_{\vec{r}}(t)= 
 \begin{cases}
	(-1)^{|\vec{r}|_1}y(t) & t_0 \le t < t_1, \\
	\quad \vdots \\
	(-1)^{|\vec{r}|_k}y(t) & t_{k-1} \le t < t_k, \\
	\quad \vdots \\
	(-1)^{|\vec{r}|_N}y(t) & t_{N-1} \le t \le t_N,
\end{cases}
\end{align}
which results from using the configuration vector $\vec{r}$ to modulate the original switching function $y(t)$. In turn, to reach the second equality of \eqref{eqc4} we exploit the Gaussian character of the noise so that the cumulant expansion truncates exactly at order two. The two contributing cumulants (see Appendix \ref{app:cumulant}),  
\begin{align}\label{c1}
\mathcal{C}^{(1)}_{\hat{o},\vec{r},\vec{r}'}(T)=2\sigma_z\int_0^T dt Y^{-,\hat o}_{\vec{r},\vec{r}'}(t)\langle B(t)\rangle=0,
\end{align}
and 
\begin{equation}\label{c2}
\begin{aligned}
&\quad {\mathcal{C}^{(2)}_{\hat{o},\vec{r},\vec{r}'}}(T) \\
&= 2I_S\int_0^T \!\! dt\!\int_0^T \!\! dt'  Y_{\vec{r},\vec{r}'}^{-,\hat o}(t) Y_{\vec{r},\vec{r}'}^{-,\hat o}(t') C^+(t,t') \\
&\quad + 4 I_S  \int_0^T \!\! dt\!\int_0^{t} \!\! dt' Y_{\vec{r},\vec{r}'}^{-,\hat o}(t) Y_{\vec{r},\vec{r}'}^{+,\hat o}(t') C^-(t,t'),\\
&\equiv 2 I_S I^+_{\vec{r},\vec{r}'}+4 I_S I^-_{\vec{r},\vec{r}'},
\end{aligned}
\end{equation}
can be compactly written in terms of the {\it effective} switching functions
\begin{equation}\label{swi}
Y_{\vec{r},\vec{r}'}^{\pm,\hat o}(t) = \frac{y_{\vec{r}}(t) \pm f_{\hat o}^z y_{\vec{r}'}(t)}{2}.
\end{equation}
Since it will be enough for our purpose to use as observables $\sigma_x$ or $\sigma_y,$ we can set $f_{\hat o}^z = -1$ and thus drop the superscript ${\hat o}$ in $Y^\pm(t)$, $I^\pm(T)$ and $v(T)$ below.  Notice that because the non-vanishing cumulant is proportional to $I_S$, the qubit's reduced dynamics is in essence a linear combination of the {\it functions} 
\begin{equation}
\label{vfunc}
v_{\hat o,\vec{r},{\vec{r}'}}(T) \equiv e^{-   I^+_{\vec{r},\vec{r}'} - 2\text{i} {\rm Im} I^-_{\vec{r},\vec{r}'}},
\end{equation}
with the specific linear coefficients depending on $\hat{O},$  $\rho_S,$ and the control. Note that $I^-$ is purely imaginary, and we write it as above to emphasize the imaginary character of the exponential associated to it. In the same vein, $I^+$ is a real quantity. The above implies that $c$ noise generates decoherence, while $q$ noise is in charge of generating a phase in our expectation values. Notice that the periodic character of the complex exponential will complicate the task of inferring information about the $q$ component of the noise. We will call this the {\it multi-value problem} and show it can be overcome via a proper analysis in Section \ref{subsec:multi-value}.

These equations encode the key ingredient of our result: the use of non-$\pi$ control in principle enables us to access functions $v_{\hat o,\vec{r},{\vec{r}'}}(T)$ which depend on integrals (i) containing {\it effective} switching functions with zeros, i.e., as if the coupling could be effectively and exactly switched off during in a portion of the evolution, and, (ii) in the case of the $q$ contribution, containing two different switching functions, allowing us to exploit their constructive/destructive interference effects for noise characterization. 

Before proceeding to show how the $v_{\hat o,\vec{r},{\vec{r}'}}(T)$ can be isolated, and how useful doing this is, it will be convenient to explore the structure of the integrals involved. 

\subsection{Properties of the effective switching functions and consequences}\label{subsec:effswi}

Let us first note that from the definition of $Y^\pm_{\vec{r},\vec{r}'}$ one has
$$|Y_{\vec{r},\vec{r}'}^\pm (t) + Y_{\vec{r},\vec{r}'}^\mp (t) | = |y_{\vec{r}}(t)| = 1,$$ and
\begin{align*}
&\quad Y_{\vec{r},\vec{r}'}^\pm (t) \times Y_{\vec{r},\vec{r}'}^\mp (t)  \\
&= [y_{\vec{r}}(t) \pm f^z_{\hat o} y_{{\vec{r}}'}(t)] [y_{\vec{r}}(t) \mp f^z_{\hat o}y_{{\vec{r}}'}(t)] /4\\
&= [y_{\vec{r}}(t)^2 -  y_{{\vec{r}}'} (t) ^2 \mp f^z_{\hat o}( y_{\vec{r}} (t) y_{{\vec{r}}'} (t) - y_{{\vec{r}}'} (t) y_{{\vec{r}}} (t))]/4=0.
\end{align*}
We dub this the {\it incompatibility condition}, as it implies that at any time  $|Y_{\vec{r},\vec{r}'}^\pm(t)| =1 $  while $|Y_{\vec{r},\vec{r}'}^\mp(t)| =0.$ That is, it cannot be that both effective switching functions vanish or are non-zero simultaneously. 

This suggests that to keep track of the effect of non-$\pi$ pulses it is enough to track the vanishing/non-vanishing character of  $Y^\pm_{\vec{r}, \vec{r}'} (t)$ in each interval. This follows from the observation that we can propose the representation 
$$Y^\pm_{\vec{r}, \vec{r}'} (t) \rightarrow Y^\pm_{(a_1,...,a_{N})} (t) = 
\begin{cases}
	a_1 y(t), & {t \in [t_{0},t_1)},  \\
	\quad \vdots \\
	a_N y(t), & {t \in [t_{N-1},t_N)},  
\end{cases}
$$
where $y(t) \in \{-1,1\}$ is a purely $\pi$-pulse generated switching function, which suggests that multiple $(\vec{r},\vec{r}')$ configurations can lead to the same $\vec{a}^{(N)} \equiv (a_1,\cdots,a_N).$ This multiplicity follows from the fact that while $\pi$ pulses define $y(t)$, they can also implement a transformation $Y^\pm_{(a_1,...,a_{N})} (t) \rightarrow Y^\pm_{(a'_1,...,a'_{N})} (t)$ with $|a_j| = |a'_j|,$ i.e., they can change the sign but not the vanishing/non-vanishing character of $Y^\pm$ in a given time interval. Concretely, a pair of $\pi$ pulses at $t_{j-1}$ and $t_j$ would enforce $a'_j = - a_j$ and $a'_i = a_i\ \forall i\neq j$. Clearly, this can also be interpreted as a change in $y(t)$, but thinking of it as a transformation of $\vec{a}^{(N)}$ is useful for our purposes.  Considering $N=3$, for example, one can achieve a {\it partial inversion} of the $Y^\pm$'s by applying extra $\pi$ pulses at $t= t_0,t_1,t_2,t_3$ so that $Y^\pm_{(a_1,0,a_3)} \rightarrow  -Y^\pm_{(a_1,0,a_3)},$ while $Y^\mp_{(0,a_2,0)} \rightarrow  +Y^\mp_{(0,a_2,0)}$. More over a {\it full inversion}, implementing $Y^\pm_{(a_1,0,a_3)} \rightarrow  -Y^\pm_{(a_1,0,a_3)}$ and $Y^\mp_{(0,a_2,0)} \rightarrow  - Y^\mp_{(0,a_2,0)}$, can be executed by applying $\pi$ pulses at $t=0,t_3$. Similar arguments follow for any $N$.

The usefulness of this ``gauge-free'' representation becomes apparent when one considers the effect of switching functions on integration domains and, in turn, of the  $v_{\hat o,\vec{r},{\vec{r}'}}(T).$  One can rewrite a given integral in terms of the $\vec{a}^{(N)}$ by making the association $I_{\vec{r},\vec{r}'}^\pm (T)  \rightarrow I_{(\vec{a},\vec{a}')}^\pm (T),$ with 
$$
I_{(\vec{a},\vec{a}')}^\pm (T) \equiv \int_0^T \!dt\! \int_0^{T'} \!dt'  Y^-_{(\vec{a})} (t)  Y^\mp_{(\vec{a}')} (t')  C^\pm(t,t'). 
$$ 
Note we have dropped the superscript $N$ to avoid cumbersome notation, since in what follows we will specialize to $N=3$.

Furthermore, let us note that  $I_{(\vec{a},\vec{a})}^+$ is invariant under a full inversion, namely $I_{(\vec{a},\vec{a})}^+ = I_{(-\vec{a},-\vec{a})}^+.$ On the other hand, the time-ordered character of $I^-$ implies that  
\begin{align*}
I_{((a_1,0,a_3), (0,a'_2,0))}^- &= I_{((0,0,a_3), (0,a'_2,0))}^- ,\\
I_{((a_1,a_2,0), (0,0,a'_3))}^- &= 0,\\
I_{((0,a_2,a_3), (-a'_1,0,0))}^- &=  I_{((0,-a_2,-a_3), (a'_1,0,0))}^- \\
& =- I_{((0,a_2,a_3), (a'_1,0,0))}^- ,\\
I_{((0,-a_2,-a_3), (-a'_1,0,0))}^- & = I_{((0,a_2,a_3), (a'_1,0,0))}^-.
\end{align*}
Notice that the above also implies that the integrals of the form $\int_{T_1}^{T_2} dt \int_{T_1}^{t}  dt'  y(t) y(t') C^-(t-t',0)$ do not contribute to the probe dynamics, and are thus in principle not learnable. This does not imply, however, that $C^-(0<t-t' < \Delta,0)$ does not contribute to the dynamics. In fact, $C^-(t-t',0)$ is in principle learnable for any $t-t' \in [0,T].$

Finally, as with the dynamical integrals, the representation $v_{\hat o,\vec{r},{\vec{r}'}}(T) \rightarrow v_{\hat o,(\vec{a},\vec{a}')}(T),$ 
where 
\begin{align*}
v_{\hat o,(\vec{a},\vec{a}')}(T) = e^{- I^+_{(\vec{a},\vec{a})}}  e^{ -2   {\rm i Im} I^-_{(\vec{a},\vec{a}')}}
\equiv  A^+_{(\vec{a})} A^-_{(\vec{a},\vec{a}')},
\end{align*}
allows us to make explicit the different ways in which $c$ and $q$ noise influence the dynamics. Namely, $c$ noise induces decay via $A^+_{(\vec{a})}$, while $q$ noise induces a phase via  $A^-_{(\vec{a},\vec{a}')}.$ 

For exposition purposes, we will often represent the former as $A^+_{(\vec{a})} = A^+{\tiny{\begin{pmatrix}
\cdot & \cdot & a_3 a'_3\\
\cdot & a_2 a'_2 & a_2 a'_3 \\
a_1 a'_1 & a_1 a'_2 & a_1 a'_3
\end{pmatrix}}},$
where the $\cdot$ makes explicit the symmetry with respect to the diagonal $(a_1a_1',a_2a_2',a_3a_3')$,  and the latter by $A^-_{(\vec{a},\vec{a}')} = A^-{\tiny{\begin{pmatrix}
\cdot & \cdot & a_3 a'_3\\
\cdot & a_2 a'_2 & a_2 a'_3 \\
a_1 a'_1 & a_1 a'_2 & a_1 a'_3
\end{pmatrix}}},$ where now the $\cdot$ makes explicit the $t \geq t'$ time ordering in the relevant integral and the matrix form facilitates keeping track of the generated equivalence  between different vector pairs of $(\vec{a},\vec{a}')$. We will say that $A^\pm (\cdot)$ is a function of the integral 
\begin{align*}
{\tiny{\begin{pmatrix}
\cdot & \cdot & a_3 a'_3\\
\cdot & a_2 a'_2 & a_2 a'_3 \\
a_1 a'_1 & a_1 a'_2 & a_1 a'_3
\end{pmatrix}}}^+ \equiv&    I^+_{(\vec{a},\vec{a}')},\\
{\tiny{\begin{pmatrix}
\cdot & \cdot & a_3 a'_3\\
\cdot & a_2 a'_2 & a_2 a'_3 \\
a_1 a'_1 & a_1 a'_2 & a_1 a'_3
\end{pmatrix}}}^- \equiv&   \text{Im} I^-_{(\vec{a},\vec{a}')}.\\
\end{align*}
Our task will be to infer the value of $I^\pm_{(\vec{a},\vec{a}')}$ for all those configurations such that $\sum_{i\leq j} |a_i a'_j| =1$ and $a_i, a'_j \in \{-1,0,1\},$ i.e., where only one of the matrix entries is non-vanishing (for $A^+$, two entries are non-zero, one of which is hidden among the dots).

\subsection{Extractable information}
\label{subsec:extractable}
To achieve this, one must first determine which among $\{v_{\hat o,(\vec{a},\vec{a}')}(T)\}_{\vec{a},\vec{a}'}$ (or linear combinations of them) can be extracted via linear combinations of expectation values. Let us first recall one is interested in expectation values of $\hat O \in\{ \sigma_x, \sigma_y\}$ after applying a set of $\sigma_y$ pulses of angles $\{\theta_i\}$ at times $\{t_i\}.$ We denote the explicit pulse dependence by $E \big[\hat{O}(t_1,...,t_{N-1},T)\big]_{\rho_S\otimes \rho_B}|_{\theta_0,...,\theta_{N}}.$ From Eq.~\eqref{VO}, one sees that each of these expectation values  can be written as a sum of terms -- each containing a specific linear combination of $v_{\hat o,(\vec{a},\vec{a}')}(T)$'s -- with coefficients of the form $\prod_{j=0}^N \cos (\frac{\theta_j  }{2})^{s_j} \sin (\frac{\theta_j  }{2})^{2-s_j}$ for $s_j \in\{ 0,1,2\}.$  This is a linear system which can be solved by cycling over an appropriate set of angles $\theta_j$ for any $N$, e.g., for $N=3$ there are 81 $ \{\theta_j\}_{j=0}^N$ configurations resulting from cycling $\theta_j \in \{ 0, \pi/2,\pi\} \, \forall \, j $, which yields the aforementioned linear combinations of interest\footnote{Note that this is a raw recipe but later we will be interested in minimizing the number of experiments, i.e., the number of configuration sets $\{\theta_j\}'s$, needed to extract a given expectation value.}. Building a linear system using $E \big[\hat{O}(t_1,...,t_{N-1},T)\big]_{\rho_S \otimes \rho_B}|_{\theta_0,...,\theta_{N}}$ for $\hat O \in\{ \sigma_x, \sigma_y\}$ and $\rho_S \in\{ \frac{I_S \pm \sigma_y}{2}, \frac{I_S \pm \sigma_x}{2}\},$ one finds that the 81 linear combinations reduce to 18 quantities that can now be accessed. Namely,
\begin{align*}
Q_1^\pm & = (A^+_{(1,0,-1)} \pm A^+_{(1,0,1)}) \!\left(A^-\tiny{\begin{pmatrix}
\cdot & \cdot & 0\\
\cdot & 0 & 1\\
0 & 0 & 0
\end{pmatrix}}  + A^-\tiny{\begin{pmatrix}
\cdot & \cdot & 0\\
\cdot & 0 & -1\\
0 & 0 & 0
\end{pmatrix}} \right)\!,\\
Q_2^\pm & = A^+_{(0,1,-1)} \left(A^-\tiny{\begin{pmatrix}
\cdot & \cdot & 0\\
\cdot & 0 & 0\\
0 & 1 & -1
\end{pmatrix}}  \pm A^-\tiny{\begin{pmatrix}
\cdot & \cdot & 0\\
\cdot & 0 & 0\\
0 & -1 & 1
\end{pmatrix}}\right),\\
Q_3^\pm & = A^+_{(0,1,1)} \left(A^-\tiny{\begin{pmatrix}
\cdot & \cdot & 0\\
\cdot & 0 & 0\\
0 & 1 & 1
\end{pmatrix}}  \pm A^-\tiny{\begin{pmatrix}
\cdot & \cdot & 0\\
\cdot & 0 & 0\\
0 & -1 &- 1
\end{pmatrix}} \right),\\
Q_4^+ & = A^+_{(0,1,0)} \left(A^-\tiny{\begin{pmatrix}
\cdot & \cdot & 0\\
\cdot & 0 & 0\\
0 & 1 & 0
\end{pmatrix}}  + A^-\tiny{\begin{pmatrix}
\cdot & \cdot & 0\\
\cdot & 0 & 0\\
0 & -1 & 0
\end{pmatrix}} \right),\\
Q_5^\pm & = A^+_{(0,0,1)} \left(A^-\tiny{\begin{pmatrix}
\cdot & \cdot & 0\\
\cdot & 0 & 1\\
0 & 0 & 1
\end{pmatrix}}  \pm A^-\tiny{\begin{pmatrix}
\cdot & \cdot & 0\\
\cdot & 0 & -1\\
0 & 0 & -1
\end{pmatrix}} \right),\\
Q_6^\pm & = A^+_{(0,0,1)} \left(A^-\tiny{\begin{pmatrix}
\cdot & \cdot & 0\\
\cdot & 0 & 1\\
0 & 0 & -1
\end{pmatrix}}  \pm A^-\tiny{\begin{pmatrix}
\cdot & \cdot & 0\\
\cdot & 0 & -1\\
0 & 0 & 1
\end{pmatrix}} \right),
\end{align*}
involving both $c$ and $q$ spectra, and those in the set  $Q|_{\pi} = \{ A^+_{(s_1, 0, 0)}, A^+_{(s_1,s_2, 0)},A^+_{(s_1,s_2,s_3)}\}$ for $s_j = \pm 1,$ involving only the $c$ noise contribution. Also note that  
$$A^-\tiny{\begin{pmatrix}
\cdot & \cdot & 0\\
\cdot & 0 & 0\\
1 & 0 & 0
\end{pmatrix}}  ,A^-\tiny{\begin{pmatrix}
\cdot & \cdot & 0\\
\cdot & 1 & 0\\
0 & 0 & 0
\end{pmatrix}}  ,A^-\tiny{\begin{pmatrix}
\cdot & \cdot & 1\\
\cdot & 0 & 0\\
0 & 0 & 0
\end{pmatrix}}  $$ are always zero and thus do not need to be learned.

\subsection{Inferrable information}\label{cla}

Inferring the values of the various integrals requires combining the accessible quantities in a non-linear way, by exploiting the fact that 
$$
A^- \tiny{\begin{pmatrix}
\cdot & \cdot & s_1\\
\cdot & s_2 & s_3\\
s_4 & s_5 & s_6
\end{pmatrix}} + A^- \tiny{\begin{pmatrix}
\cdot & \cdot & -s_1\\
\cdot & -s_2 & -s_3\\
-s_4 & -s_5 & -s_6
\end{pmatrix}}  \!= 2 \cos( {2\tiny{\begin{pmatrix}
\cdot & \cdot & s_1\\
\cdot & s_2 & s_3\\
s_4 & s_5 & s_6
\end{pmatrix}}^-} ) 
$$
and 
$$
A^-\! \tiny{\begin{pmatrix}
\cdot & \cdot & s_1\\
\cdot & s_2 & s_3\\
s_4 & s_5 & s_6
\end{pmatrix}} - A^- \!\tiny{\begin{pmatrix}
\cdot & \cdot & -s_1\\
\cdot & -s_2 & -s_3\\
-s_4 & -s_5 & -s_6
\end{pmatrix}}\!  =\! -2 \normalsize{\text{i}} \sin (2{\tiny{\begin{pmatrix}
\cdot & \cdot & s_1\\
\cdot & s_2 & s_3\\
s_4 & s_5 & s_6
\end{pmatrix}\!}^-  }).
$$

With this in hand, one can for example start by combining $Q_5^\pm$ and $Q_6^\pm$ to achieve
\begin{align*}
\frac{Q_5^+ \times Q_6^+ -  Q_5^- \times Q_6^-}{4} &=  A^+_{(0,0,{\sqrt{2}})} \cos(2\tiny{\begin{pmatrix}
\cdot & \cdot & 0\\
\cdot & 0 & 0\\
0 & 0 & 2
\end{pmatrix}}^-)\\
\frac{Q_5^+ \times Q_6^+ +  Q_5^- \times Q_6^-}{4} &=  A^+_{(0,0,{\sqrt{2}})} \cos(2\tiny{\begin{pmatrix}
\cdot & \cdot & 0\\
\cdot & 0 & 2\\
0 & 0 & 0
\end{pmatrix}}^-)\\
\frac{Q_5^+ \times Q_6^- + Q_5^- \times Q_6^+}{4 \normalsize{\text{i}} } &=  -A^+_{(0,0,\sqrt{2})} \sin(2\tiny{\begin{pmatrix}
\cdot & \cdot & 0\\
\cdot & 0 & 2\\
0 & 0 & 0
\end{pmatrix}}^-)\\
\frac{Q_5^+ \times Q_6^- -  Q_5^- \times Q_6^+}{4 \normalsize{\text{i}}} &= A^+_{(0,0,\sqrt{2})} \sin(2\tiny{\begin{pmatrix}
\cdot & \cdot & 0\\
\cdot & 0 & 0\\
0 & 0 & 2
\end{pmatrix}}^-).
\end{align*}
This provides the knowledge of $\tiny{\begin{pmatrix}
\cdot & \cdot & 1\\
\cdot & 0 & 0\\
0 & 0 & 0
\end{pmatrix}}^+, \tiny{\begin{pmatrix}
\cdot & \cdot & 0\\
\cdot & 0 & 1\\
0 & 0 & 0
\end{pmatrix}}^- \,\, ({\rm mod} (2 \pi))$ and $\tiny{\begin{pmatrix}
\cdot & \cdot & 0\\
\cdot & 0 & 0\\
0 & 0 & 1
\end{pmatrix}}^- ({\rm mod} (2 \pi)),$ which can in turn be leveraged in $Q_1^\pm$ to infer $A^+_{(1,0,-1)} \pm A^+_{(1,0,1)}.$ Further, since one has that 
\begin{align*} 
A^+_{(1,0,-1)} + A^+_{(1,0,1)}= 2 A^+_{(1,0,0)}  A^+_{(0,0,1)}   \cosh (2 \tiny{\begin{pmatrix}
\cdot & \cdot & 0\\
\cdot & 0 & 0\\
0 & 0 & 1
\end{pmatrix}}^+),\\
A^+_{(1,0,-1)} - A^+_{(1,0,1)}= 2 A^+_{(1,0,0)}  A^+_{(0,0,1)}   \sinh (2 \tiny{\begin{pmatrix}
\cdot & \cdot & 0\\
\cdot & 0 & 0\\
0 & 0 & 1
\end{pmatrix}}^+),
\end{align*}
and $\tiny{\begin{pmatrix}
\cdot & \cdot & 1\\
\cdot & 0 & 0\\
0 & 0 & 0
\end{pmatrix}}^+$ and $\tiny{\begin{pmatrix}
\cdot & \cdot & 0\\
\cdot & 0 & 0\\
1 & 0 & 0
\end{pmatrix}}^+$ are known, one can then also infer $\tiny{\begin{pmatrix}
\cdot & \cdot & 0\\
\cdot & 0 & 0\\
0 & 0 & 1
\end{pmatrix}}^+.$

Then, using $Q_2^\pm$ and $Q_3^\pm,$ one finds that 
\begin{align*}
\frac{Q_2^+ \times Q_3^+ -  Q_2^- \times Q_3^-}{4  A^+_{(0,0,\sqrt{2})}} &= A^+_{(0,\sqrt{2},0)} \cos(2\tiny{\begin{pmatrix}
\cdot & \cdot & 0\\
\cdot & 0 & 0\\
0 & 0 & 2
\end{pmatrix}}^-),\\
\frac{Q_2^+ \times Q_3^+ +  Q_2^- \times Q_3^-}{4 A^+_{(0,0,\sqrt{2})}} &=  A^+_{(0,\sqrt{2},0)}\cos(2\tiny{\begin{pmatrix}
\cdot & \cdot & 0\\
\cdot & 0 & 0\\
0 & 2 & 0
\end{pmatrix}}^-),\\
\frac{Q_2^+ \times Q_3^- + Q_2^- \times Q_3^+}{-4 {\text{i}} A^+_{(0,0,\sqrt{2})}} &=  A^+_{(0,\sqrt{2},0)} \sin(2\tiny{\begin{pmatrix}
\cdot & \cdot & 0\\
\cdot & 0 & 0\\
0 & 2 & 0
\end{pmatrix}}^-),\\
\frac{Q_2^+ \times Q_3^- -  Q_2^- \times Q_3^+}{-4 {\text{i}} A^+_{(0,0,\sqrt{2})}} &= A^+_{(0,\sqrt{2},0)} \sin(2\tiny{\begin{pmatrix}
\cdot & \cdot & 0\\
\cdot & 0 & 0\\
0 & 0 & 2
\end{pmatrix}}^-),
\end{align*}
which not only makes $Q_4^+$ superfluous, but gives us access to the integrals $\tiny{\begin{pmatrix}
\cdot & \cdot & 0\\
\cdot & 0 & 0\\
0 & 1 & 0
\end{pmatrix}}^- ({\rm mod} (2 \pi))$ and consequently to $A^+_{(0,1, 0)}$. The final step is to realize that 
$$
\frac{A^+_{( 1,1,0)}}{A^+_{( 1,-1,0)}} = A^+ {\tiny{\begin{pmatrix}
\cdot & \cdot & 0\\
\cdot & 0 & 0\\
0 & 2 & 0 
\end{pmatrix}}}   \textrm{    and    }  \frac{A^+_{( 0,1,1)}}{A^+_{( 0,1,-1)}} = A^+ {\tiny{\begin{pmatrix}
\cdot & \cdot & 0\\
\cdot & 0 & 2\\
0 & 0 & 0 
\end{pmatrix}}}, 
$$ 
which completes the picture. That is, one can infer the value (up to a multiple of $ 2\pi $ in the quantum case) of integrals of the form 
\begin{align*}
\mathcal{I}^\pm(\vec{T})\equiv\int_{{T_3}}^{{T_4}} dt \int _{{T_1}}^{{T_2}} dt' Y^-(t) Y^\mp(t') C^{\pm}(t-t',0) 
\end{align*}
where $T_{i+1}-T_{i} \geq \Delta.$ and ${T_3 \geq T_2 }$.

\subsubsection{The multivalue problem}\label{subsub:multi}
An important caveat in the above argument is that, as highlighted, the $c$ and $q$ spectra appear in two different ways. Importantly, the latter appears within a periodic function. This is specially significant in our scenario as we want to be able to study the strong coupling regime. Thus, it need not be that the relevant integrals have values within the first period of $\sin$ or $\cos$ functions. In other words, one can only estimate  $I^-_{(\vec{a},\vec{a}')}$ modulo $2\pi$. If the noise is weak such that one can assume $I^-_{(\vec{a},\vec{a}')} < 2 \pi,$ this is not a problem. In the general strong coupling regime this represents a considerable obstacle to the accurate characterization of the $q$ component of the noise. For now, we shall assume that the problem can be solved, and will indeed provide a solution later on when all the necessary tools have been deployed.

\subsection{The power of control and stationarity considerations}

Stationarity has a considerable effect in what sort of integrals can be learned. To explore it let us first note that 
\begin{align*}
&\int_{T_3}^{T_4} dt \int _{T_1}^{T_2} dt' y(t) y(t') C^{\pm}(t-t',0) \\
=& \int_{T_3-d}^{T_4-d } dt \int _{T_1-d }^{T_2-d } dt'  {y}'(t) {y}'(t') C^{\pm}(t-t',0)
\end{align*}
holds for any $d$, provided that ${y}'(t-d) {y}'(t'- d) = y(t) y(t').$ This implies that $\mathcal{I}^\pm(\vec{T})$ can be reduced to an integral like
\begin{align}\label{eqg1}
   \mathcal{I}^\pm(t_1, t_2,T) \equiv \int_{t_2}^T dt \int_0^{t_1} dt' Y^-(t) Y^\mp(t') C^\pm (t-t',0)  
\end{align}
for $t_2\geq t_1$ provided suitable $\pi$-pulse configurations have been chosen. Given fixed $t_1$, however, it would seem that the value of $\mathcal{I}^\pm(t_1, t_2 < t_1,T)$ -- thought of as a function of $t_2$ --  cannot be directly accessed. This apparent limitation can be overcome by combining information from different experimental configurations.

\begin{figure*}[th]
 \begin{tabular}{@{}p{0.46\linewidth}p{0.02\linewidth}p{0.485\linewidth}@{}}
   \subfigimg[width=\linewidth]{(a)}{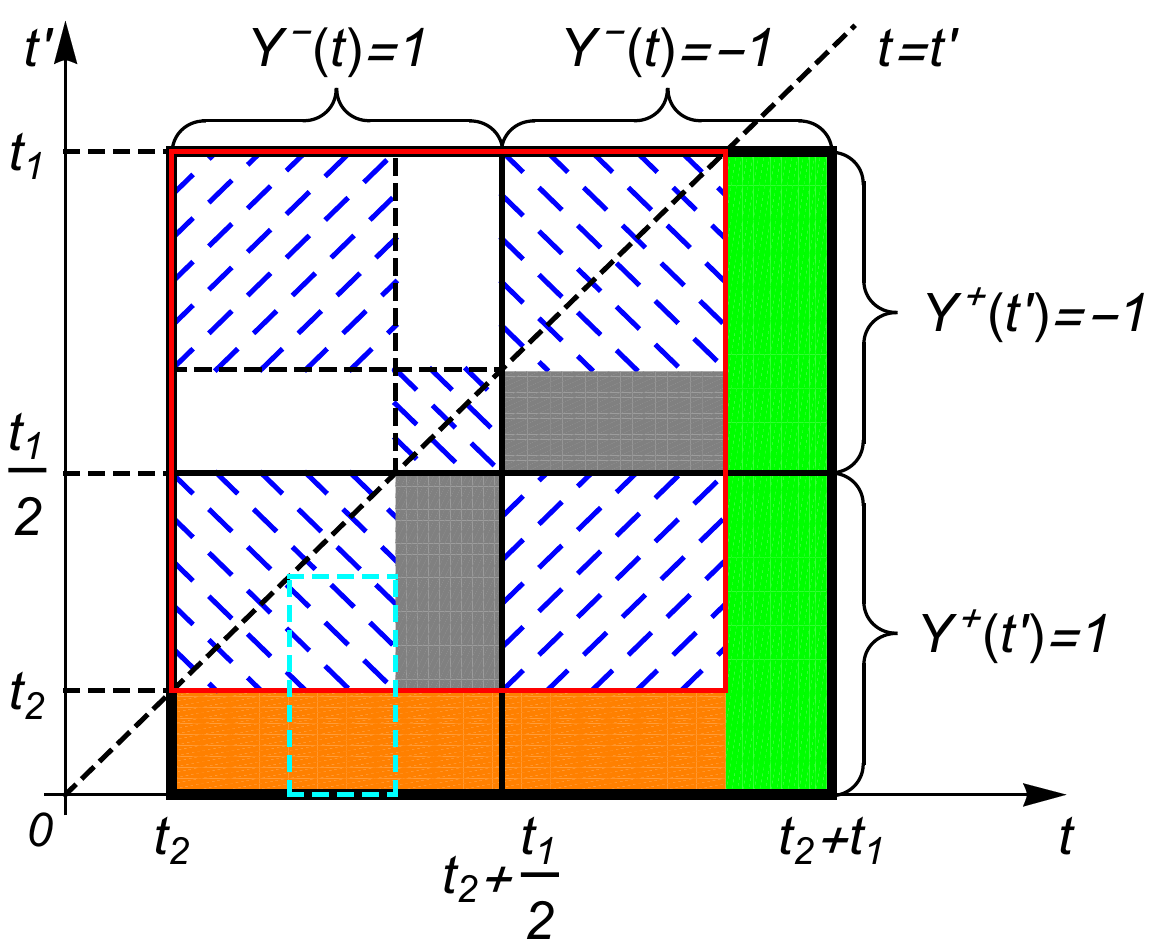} & &
   \subfigimg[width=\linewidth]{(b)}{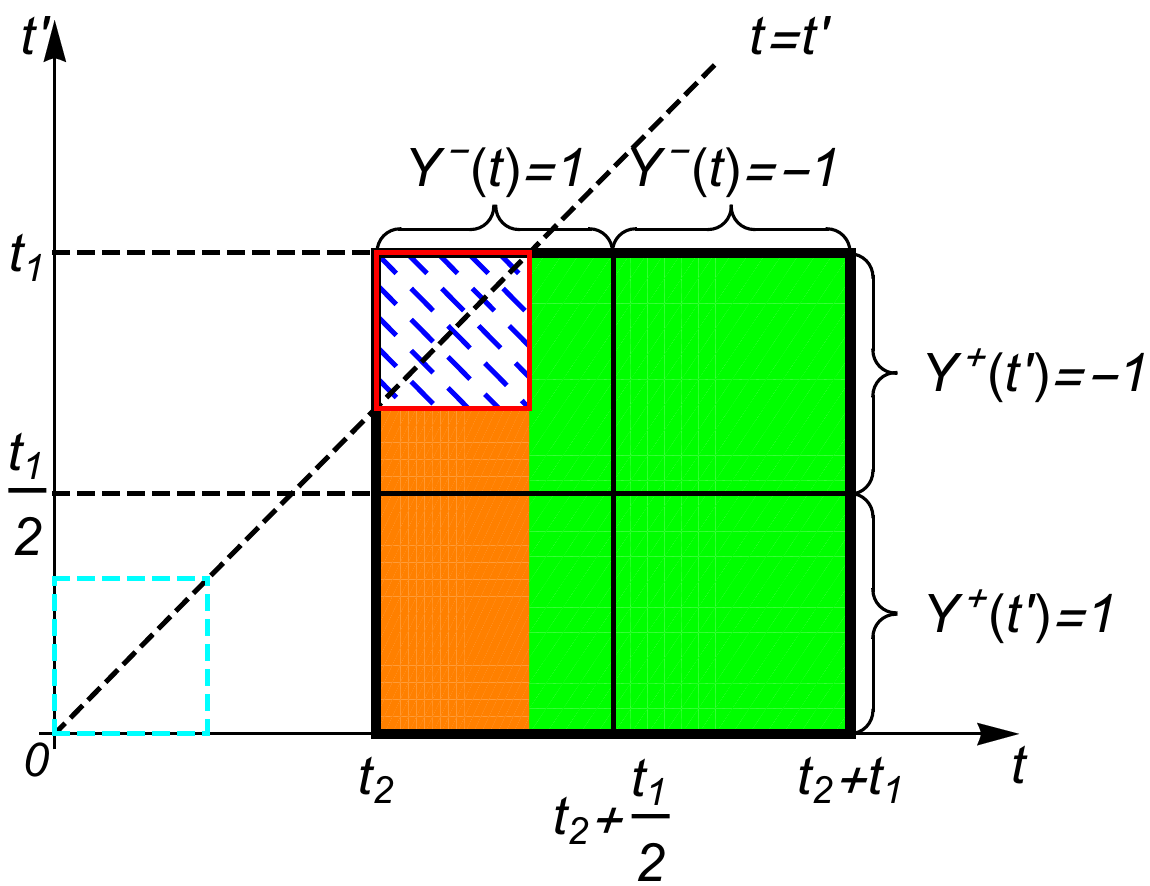} 
  \end{tabular}
\vspace{-1em}
\caption{Exemplified schematic diagrams to infer the value of $\mathcal{I}^\pm(t_1, t_2<t_1,T)$. The lines $t=t_2+t_1/2$ and $t'=t_1/2$ correspond to the effective switching functions' $\pi$ pulse to realize a Hahn Echo filter. The orange and green regions can each be accessed by individual experiments, and the key remained is $\mathcal{I}^\pm(t_2,t_1,t_2,t_1)$, the red region. (a) $0<t_2<t_1/2$. For $\mathcal{I}^-$, the integrals on the blue dashed regions sum to zero due to antisymmetry of the $q$-noise. Integrals on the two gray regions are equal by stationarity, and hence the red region integral value reduces to four times the integral value of the cyan-dashed region, $\mathcal{I}^\pm(t_1/2-t_2,t_1/2-t_2,t_1/2)$, which can be obtained by an experiment with a non-$\pi$ pulse at $t_1/2-t_2$. For $\mathcal{I}^+$, the red region integral reduces to the blue dashed regions and can be experimentally accessed individually using stationarity. (b) $t_1/2\leq t_2<t_1$. The red region is zero for $\mathcal{I}^-$, while for the $c$-spectrum equivalent to the cyan square $\mathcal{I}^+(0,t_1-t_2,0,t_1-t_2)$, which can be accessed by a single experiment.}
\label{fig:deadtime}
\end{figure*}

To see this, let us consider the situation where each $t_i$ and $|t_1-t_2|$ is zero or greater than $\Delta$. Then, one can access the value of $\mathcal{I}^\pm(t_1, t_2 < t_1,T)$ using the scheme in 
Fig.~\ref{fig:deadtime}, where the $\pi$ pulses form a Hahn Echo filter while other pulse sequences are also allowed. Notice that depending on the $c$/$q$ spectra, one must be aware of the incompatibility condition. One starts then by implementing an experiment granting access to $\mathcal{I}^\pm(t_a,t_b,t_c)$ with $t_a = t_b =t_1$ and $t_c= T$, which corresponds to the green shaded region of the integral. Additionally, non-$\pi$ pulses at $t_a = t_b=t_2$ and $t_c= t_1$ give us access to $\mathcal{I}^\pm(t_2,t_2,t_1)$, the orange region. Then importantly, one can perform a carefully-designed experiment to indirectly infer the value of $\mathcal{I}^\pm(t_2,t_1,t_2,t_1)$, the red region, with two such examples provided in Fig. \ref{fig:deadtime}. Finally, adding up the integral values of the three colored regions, one obtains the value of $\mathcal{I}^\pm(t_1, t_2<t_1,T)$ as desired.

{\it Reaching the control limit.-} Control constraints naturally impose a limit to what information can be learnt. In the presence of a minimum switching time $\Delta,$ the above argument shows one can access information about $\mathcal{I}^\pm(t_1, t_2,T)$ for $ \Delta \leq t_1, t_2 \leq  T,$ with the difference between any two pulse timings being $0$ or not smaller than $\Delta.$ In particular, one is able to infer the value of $\mathcal{I}^\pm(\Delta, k \Delta, (k+1) \Delta)$ for $k>0$. Notice that this can be done in a history-independent way by applying our basic three-interval setup using $t_1 = \Delta, t_2= k \Delta, T= (k+1) \Delta.$ The argument can obviously be extended to any grid size larger than $\Delta$, provided it is an integer multiple of $\delta.$ 

Control can take us further, if one is willing to pay the cost. The previous set up required a small number of experiments to infer $\mathcal{I}^\pm(t_1 =\Delta, t_2=k \Delta, T= (k+1) \Delta),$ or any appropriate $t_i$ for that matter. Using experiments scaling with $\Delta/\delta,$ however, one can infer values of $\mathcal{I}^\pm(t_1, t_2,T)$ limited by the time solution $\delta.$

\begin{figure}
\includegraphics[scale=0.35]{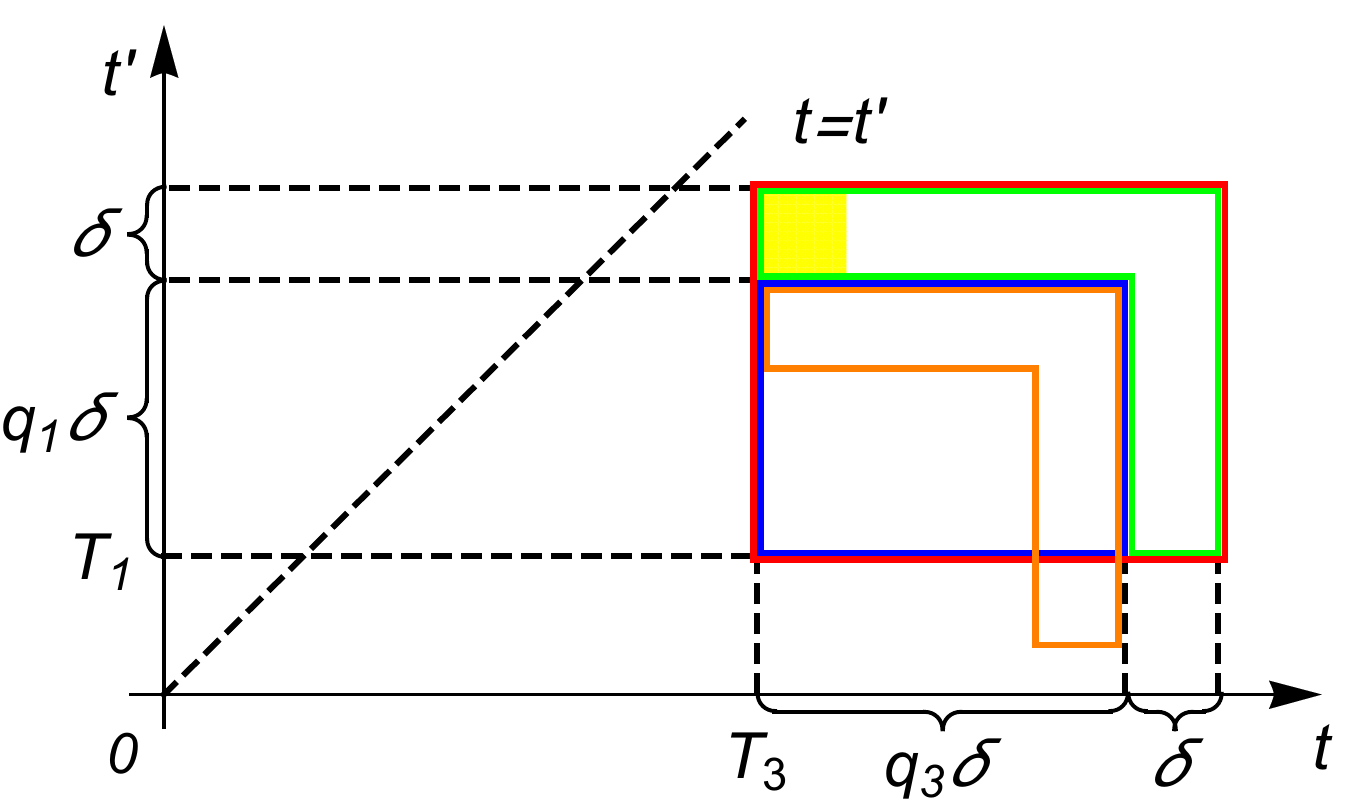}
\vspace{-1em}
\caption{A schematic plot illustrating the procedures to infer the value of $\mathcal{I}_0^\pm(T_1+q_1 \delta,T_1+(q_1+1) \delta, T_3,T_3+ \delta)$, the yellow square. Starting from the integral values of the blue and red squares, their difference is the value of the green strip. Similarly the orange strip is accessed. Finally the difference between the orange and green strips is the yellow square.}
\label{fig:strip}
\end{figure}

Starting form the previous result, consider now the integrals 
\begin{align*}
&\mathcal{I}_0^\pm(T_1,T_1+ q_1  \delta,T_3,T_3+q_3\delta),\\
&\mathcal{I}_0^\pm(T_1,T_1+ (q_1+1)  \delta,T_3,T_3+(q_3+1)\delta),
\end{align*}
with $q_i \delta \geq \Delta$ and the $0$ subscript denoting the scenario where $y(t) = y'(t) = 1$ (Thus $\mathcal{I}_0^-(T_1,T_2,T_1,T_2)=0$), corresponding to the blue and red squares in Fig. \ref{fig:strip}.   
By subtracting them, one can infer the value of an ``angle" strip (the green enclosed region). One can then build another such ``angle" strip (the orange region) with $T'_1 = T_1 - \delta, q'_1= q_1$ and $T'_3 = T_3, q'_3=q_3 -1$, so that the difference between the ``angles'' is the yellow square $\mathcal{I}_0^\pm(T_1+q_1 \delta,T_1+(q_1+1) \delta, T_3,T_3+ \delta),$ for $q_1 \geq \lceil \Delta/\delta\rceil,$ i.e., the average of $C^\pm(t-t',0)$ over a $\delta \times \delta$ integration domain. Taking this to the extreme, one can in principle infer integrals of the form $\mathcal{I}_0^\pm(0,\delta,{q} \delta, ({q}+1) \delta  ),$ for any positive integer $q\geq \lceil \Delta/\delta\rceil$. We stress that the lower bound on $q$ comes from the minimum switching time, as it also implies a minimum measurement time $T.$ Notice, that if the lower bound on the measurement time is not present, then there is no lower bound on $q$ above. 

Since any integral $\mathcal{I}_0^\pm(t_1,t_2,T)$ can be reconstructed by an appropriate linear combination of 
$\{ \mathcal{I}_0^\pm(0,\delta,{q} \delta, ({q}+1) \delta  )\},$ this represents the most detailed information that can be inferred using the declared control constraints. While access to such detailed information can undoubtedly be very useful, its acquisition may not be practical. This is specially true if in addition to counting raw resources, e.g,. number of necessary experimental configurations, one also considers extra experimental limitations, such as measurement and finite-sampling errors. Thus, here we will use this full-access reconstruction as a benchmark to our  more cost-effective, but less detailed, reconstruction protocols, and will leave its detailed analysis for future work.

\subsection{Overcoming limitations}

The result in the previous section allows us to overcome the limitations associated to dephasing-preserving control (assuming for now the multivalue problem is solved).

{\it (i) } The $q$ component of the noise can be sampled, in a manner that is {\it weakly} dependent on the $c$ component. Consider, for example, the ability to access 
\begin{align*}
A^+_{(0,0,1)} \left(A^-\tiny{\begin{pmatrix}
\cdot & \cdot & 0\\
\cdot & 0 & 1\\
0 & 0 & 1
\end{pmatrix}}  \pm A^-\tiny{\begin{pmatrix}
\cdot & \cdot & 0\\
\cdot & 0 & -1\\
0 & 0 & -1
\end{pmatrix}} \right)
\end{align*}
via a linear combination of expectation values. Notice that the $c$ contribution of the noise is restricted to the domain $[t_2,T] \times [t_2,T],$ and thus the $q$ component does not depend on the history of the contribution before $t_2$. This is important, as one can now make $t_2$ and $T$ in principle arbitrarily large, while maintaining $|T-t_2|$ fixed, i.e., maintaining the ``damping'' of the signal due to the $c$ noise contribution restricted. 

This can be taken farther by assuming that control for $t \in [t_2,T]$ is such that  $\int_{t_2}^T  dt \int_{t_2}^T  dt'  y(t) y(t') C^+(t-t',0) \rightarrow 0.$ While this limit is not physically achievable, in principle a sufficiently powerful error suppression mechanism like dynamical decoupling can make the integral small. A similar argument has been considered in \cite{shen2023}. We highlight, however, that the validity of this simplification depends on $C^+(t-t',0)$ being well matched to the mechanism being used, e.g., DD requires that the noise is mostly low frequency to be effective. Without a prior knowledge of  $C^+(t-t',0)$ or relevant assumptions in place, this is an oversimplification which will eventually reduce us to seek qualitatively correct~\cite{shen2023} -- rather than quantitatively accurate -- estimations of $C^\pm(t-t',0).$ We will not make such approximations in this paper to maintain generality and access to the full information provided by the probe's dynamics.

{\it (ii)}  There is in principle no lower bound to the frequency that can be sampled.  Combining our ability to infer the different integrals, one has now the ability to infer the value of $\mathcal{I}^\pm(t_1,t_2,T)$ as \eqref{eqg1}, which allows us to obtain knowledge of  $C^\pm(\tau,0)$ for $\tau \geq \Delta$, without significant concerns for the history of dynamics, as was the case in the dephasing-preserving control scenario. This is a direct consequence of the weak dependence of the accessible quantities, i.e., one can access simple quantities involving the $q$/$c$ component with only a weak dependence on the $c$/$q$ component. We stress that weak does not mean non-existent, and the codependence will impact -- if in a relatively minor and controllable way -- our ability to accurately extract noise correlation information, as we will see in the next section.

\subsection{Specific measurement equations}\label{subsec:equs}

The deduction in Sec. \ref{cla} demonstrates the existence of a systematic method to infer the knowledge of the integral values from the observed expectation values. However, the specific initial states and observables are not listed and the equations therein might not always be the most efficient, if considering to minimize the number of experimental configuration sets $(\rho_S,\hat{O},N,\theta_0,...,\theta_N)$ needed. Hence, here we list alternative full equations showing how to extract information efficiently. In this subsection, $\theta_0$ and $\theta_N$ are nonexistent and pulses are only applied at $t_1,...,t_{N-1}$, denote $k_i\in \mathbb{N}$, $\rho_S\otimes \rho_B$ is simplified as $\ket{\pm}_\alpha$ where $\rho_S=\frac{I_S\pm\sigma_\alpha}{2}$, and we take $N=3$ unless otherwise specified. {We also employ stationarity to reduce the number of integral values of interest. For example, having access to $A^+(1,0,0)$ suffices to infer the values of $A^+(0,1,0)$ and $A^+(0,0,1)$.} 

(i) The elements in $Q|_\pi$ can be accessed from
\begin{equation}\label{eqb4}
\quad E \Big[\hat{Y}(t_1,t_2,T)\Big]_{|+\rangle_y}\Big|_{k_1\pi,k_2\pi}\!\!=A^+_{(1,(-1)^{k_1},(-1)^{k_1+k_2})}.
\end{equation}
Also, using stationarity, the value of $A^+_{(0,0,1)}$ equals to the result of an $N=1$ experiment of
\begin{align}\label{eqx01}
E \Big[\hat{Y}(T-t_2)\Big]_{|+\rangle_y},
\end{align}
which will be used later for other purposes.

(ii) To access $\tiny{\begin{pmatrix}
\cdot & \cdot & 0\\
\cdot & 0 & 0\\
0 & 1 & 0
\end{pmatrix}}^+$, $N=2$ intervals suffice and we have
\begin{align}
&\quad E \Big[\hat{Y}(t_1,T)\Big]_{|+\rangle_y}\Big|_{k_1\pi}\nonumber\\
&=A^+_{(1,0)}A^+_{(0,1)}\exp \Big((-1)^{k_1+1}{\scriptsize{\begin{pmatrix}
\cdot & 0 \\
0 & 1 
\end{pmatrix}}}^+\Big).\label{eqx27}
\end{align}
In practice, one can first consume a small number of state copies to roughly estimate which one of $E(\hat Y)|_0,E(\hat Y)|_\pi$ in \eqref{eqx27} is larger, and then only employ the corresponding equation. Through \eqref{eqb4}, one knows the value of $A^+_{(1,0)}A^+_{(0,1)}$, and finally can infer ${\tiny{\begin{pmatrix}
\cdot & 0 \\
0 & 1 
\end{pmatrix}}}^+$.

(iii) To access $\tiny{\begin{pmatrix}
\cdot & \cdot & 0\\
\cdot & 0 & 0\\
0 & 0 & 1
\end{pmatrix}}^+$ such that the low-frequency information of $c$-component can be extracted, we first denote 
\begin{align}\label{eqx11}
\mathcal{A}&\equiv E\Big[\hat{Y}\Big]_{|+\rangle_y}\Big|_{\frac{\pi}{2},\frac{\pi}{2}}-E \Big[\hat{Y}\Big]_{|+\rangle_y}\Big|_{\frac{\pi}{2},-\frac{\pi}{2}}\\
&+E \Big[\hat{Y}\Big]_{|+\rangle_y}\Big|_{-\frac{\pi}{2},-\frac{\pi}{2}}-E \Big[\hat{Y}\Big]_{|+\rangle_y}\Big|_{-\frac{\pi}{2},\frac{\pi}{2}},\notag 
\end{align}
\begin{align}
\mathcal{B}&\equiv E \Big[\hat{X}\Big]_{|-\rangle_x}\Big|_{\frac{\pi}{2},\frac{\pi}{2}}+E \Big[\hat{X}\Big]_{|+\rangle_x}\Big|_{-\frac{\pi}{2},\frac{\pi}{2}}.
\label{eqa1}
\end{align}
Then after collecting the values of $\mathcal{A},\mathcal{B}$, we can deduce $\tiny{\begin{pmatrix}
\cdot & \cdot & 0\\
\cdot & 0 & 0\\
0 & 0 & 1
\end{pmatrix}}^+$ from
\begin{align}
{\rm{sgn}}(\frac{\mathcal{A}}{\mathcal{B}})\frac{|\mathcal{A}|}{\sqrt{4\mathcal{B}^2-\mathcal{A}^2}}=&\sinh\Big({\scriptsize{\begin{pmatrix}
\cdot & \cdot & 0\\
\cdot & 0 & 0\\
0 & 0 & 1
\end{pmatrix}}^+}\Big).\label{eqa6}
\end{align}

The specifics about how to exploit Eq. \eqref{eqa6} efficiently in practice are presented in Appendix \ref{sec:prac}.

(iv) To access $\tiny{\begin{pmatrix}
\cdot & \cdot & 0\\
\cdot & 0 & 1\\
0 & 0 & \pm 1
\end{pmatrix}}^-$, we notice that
\begin{align}
&\quad E \Big[\hat{X}(t_1,t_2,T)\Big]_{|+\rangle_z}\Big|_{k_1\pi,\frac{\pi}{2}}\notag\\
&=(-1)^{k_1}A^+_{(0,0,1)}\cos\Big(2{\tiny{\begin{pmatrix}
\cdot & \cdot & 0\\
\cdot & 0 & (-1)^{k_1}\\
0 & 0 & 1
\end{pmatrix}}}^-\Big),\label{eqb14}
\end{align}
\begin{align}
&\quad  E \Big[\hat{Y}(t_1,t_2,T)\Big]_{|+\rangle_z}\Big|_{k_1\pi,\frac{\pi}{2}}\notag\\
&=(-1)^{k_1}A^+_{(0,0,1)}\sin\Big(2{\tiny{\begin{pmatrix}
\cdot & \cdot & 0\\
\cdot & 0 & (-1)^{k_1}\\
0 & 0 & 1
\end{pmatrix}}}^-\Big).\label{eqb14a}
\end{align}
Note that Eq. \eqref{eqx01} gives the value of $A^+_{(0,0,1)}$, which after input to Eq. \eqref{eqb14a} leads to $\tiny{\begin{pmatrix}
\cdot & \cdot & 0\\
\cdot & 0 & 1\\
0 & 0 & \pm 1
\end{pmatrix}}^-$.

(v) To access $\tiny{\begin{pmatrix}
\cdot & \cdot & 0\\
\cdot & 0 & 0\\
0 & 0 & 1
\end{pmatrix}}^-$, one can employ Eqs. \eqref{eqb14} and \eqref{eqb14a} to achieve
\begin{align}
&\quad E \Big[\hat{X}(t_1,t_2,T)\Big]_{|+\rangle_z}\Big|_{0,\frac{\pi}{2}}E \Big[\hat{Y}(t_1,t_2,T)\Big]_{|+\rangle_z}\Big|_{\pi,\frac{\pi}{2}}\nonumber\\
&+E \Big[\hat{Y}(t_1,t_2,T)\Big]_{|+\rangle_z}\Big|_{0,\frac{\pi}{2}}E \Big[\hat{X}(t_1,t_2,T)\Big]_{|+\rangle_z}\Big|_{\pi,\frac{\pi}{2}}\nonumber\\
=&-A^+_{(0,0,\sqrt{2})}\sin \Big(4{{\tiny{\begin{pmatrix}
\cdot & \cdot & 0\\
\cdot & 0 & 0\\
0 & 0 & 1
\end{pmatrix}}}}^-\Big),\label{eqx02}
\end{align}
\begin{align}
&\quad E \Big[\hat{X}(t_1,t_2,T)\Big]_{|+\rangle_z}\Big|_{0,\frac{\pi}{2}}E \Big[\hat{X}(t_1,t_2,T)\Big]_{|+\rangle_z}\Big|_{\pi,\frac{\pi}{2}}\nonumber\\
&-E \Big[\hat{Y}(t_1,t_2,T)\Big]_{|+\rangle_z}\Big|_{0,\frac{\pi}{2}}E \Big[\hat{Y}(t_1,t_2,T)\Big]_{|+\rangle_z}\Big|_{\pi,\frac{\pi}{2}}\nonumber\\
=&-A^+_{(0,0,\sqrt{2})}\cos \Big(4{{\tiny{\begin{pmatrix}
\cdot & \cdot & 0\\
\cdot & 0 & 0\\
0 & 0 & 1
\end{pmatrix}}}}^-\Big).\label{eqx03}
\end{align}

Close to the end of Sec. \ref{subsec:dynamics}, we have pointed out that $q$ noise introduces a phase to the observable expectation value and induces no decoherence, in sharp contrast with $c$ noise. Despite this, to estimate the low-frequency information of $q$ noise, it is still necessary to generate zero-value filter functions to avoid decoherence resulting from $c$ noise, considering that $c$ noise always accompanies the appearance of $q$ noise (while the contrary is not true), as exemplified by $A^+_{(0,0,1)}$ in Eq. \eqref{eqb14}. The advantage of Eqs. \eqref{eqx02} and \eqref{eqx03} is that, when $T$ is forced to be large in order to probe the long-time correlation of $q$ noise, the RHSs do not decohere, so long as $T-t_2$ is not overlarge.

Both (iv) and (v) require to implement $\arcsin(\cdot)$, invoking the multivalue problem, the definition and solution of which are detailed in Sec. \ref{subsub:multi} and Sec. \ref{subsec:multi-value}. Eq. \eqref{eqb14} will also be employed in the solution, while consuming only a small number of state copies to roughly determine the sign of $\cos(2{\tiny{\begin{pmatrix}
\cdot & \cdot & 0\\
\cdot & 0 & (-1)^{k_1}\\
0 & 0 & 1
\end{pmatrix}}}^-)$.

\section{Frequency-domain analysis}\label{sec:freq}

The previous sections equipped us with access to the fundamental integral values of the correlation functions in the time domain. As is common in signal processing, here we move to the frequency domain to complete our analysis. We expect to gain two things from this: (i) further insight on the noise structure, and (ii) a way to achieve such insight at a reduced cost, i.e., without requiring information about the bath correlations in every integration domain.  

\subsection{From time to frequency}\label{subsec:fre1}

The key quantity in the frequency domain analysis of stationary noise processes is the  power spectrum $S(\omega),$ which follows from the Fourier transform $\mathcal{F}[\cdot](\vec{\omega})$ of the cumulant, i.e.,
\begin{eqnarray}
\nonumber {\mathcal{F}[C^{(2)}( B(t), B(t'))] (\omega_1, \omega_2)}\!\! &=& \!\!\!\!\!  \int_{\mathbb{R}^2} \!\!\!\! d \vec t e^{- {\rm i} \vec{\omega}\cdot \vec t}  C^{(2)} (B(t), B(t')) \\
& \equiv &\!\! {2\pi} \delta(\omega_1 + \omega_2) S(\omega_1).
\end{eqnarray}
In the same way one can write the symmetric and anti-symmetric versions of the correlation function, i.e., $C^{\pm} (t-t',0),$ and the $c$/$q$ power spectra are given by 
$$ S^{\pm} (\omega) = \mathcal{F} [C^\pm(\tau,0)](\omega),$$ where we note that the $c$ spectrum $S^+(\omega)$ is symmetric about $\omega=0$ and non-negative, while the $q$ spectrum $S^-(\omega)$ is antisymmetric and can be negative. In this language, noticing that $y(t)$ can be built to have a completely different $\pi$-pulse generated structure in the intervals $[0,t_1]$ and $[t_2,T]$ and choosing  $T = t_2 + t_1,$ one can rewrite
\begin{align}
\nonumber &\quad \mathcal{I}^\pm(t_1,t_2,T)\\
\nonumber &= \int_{t_2}^{t_2 + t_1} dt \int_{0}^{t_1} dt'  y(t) y'(t') C^{\pm}(t-t',0),\\
\nonumber &= \frac{1}{2\pi}\int_{-\infty}^{\infty} d\omega e^{{\rm i}\omega t_2} F(\omega, t_1) F'(-\omega, t_1) S^\pm (\omega)\\
&=\left\{\!\!\!\!\!\!\begin{array}{rl}
&\frac{1}{2\pi}\int_{-\infty}^{\infty}\!\! d\omega \, {\rm Re} [e^{{\rm i}\omega t_2} F(\omega,  t_1) F'(-\omega, t_1)] S^+(\omega)\vspace{1ex}\\
&\frac{{\rm i}}{2\pi} \int_{-\infty}^{\infty} \!\! d\omega \, {\rm Im} [e^{{\rm i}\omega t_2} F(\omega,  t_1) F'(-\omega, t_1)] S^-(\omega)
\end{array}\!\!,\right.\label{eqa10}
\end{align}
where $F'(\omega,t_1) \equiv \int_0^{ t_1} d\tau\ e^{{\rm i}\omega \tau} y'(\tau)$ and $F(\omega, t_1)\equiv \int_0^{t_1} d\tau\ e^{{\rm i}\omega \tau} y(\tau)$ are the  filter functions associated to the intervals $[0,t_1]$ and $[t_2,T=t_2+t_1]$, respectively. Eq. \eqref{eqa10} is the cornerstone of this paper, as it opens multiple avenues for improving existing frequency-domain QNS protocols and overcoming some of their most notable limitations. 

Concretely, we will showcase a method using the full Fourier machinery and outputting a fine-grained sample of the spectra, with a resolution determined by the experimental limitations.

\subsection{Fourier-based reconstruction}\label{subsec:fouriertrans}

This algorithm stems from the observation that when (i) $T-t_2 = t_1$ and (ii) combining results from when the sequences in intervals $t \in [t_2,T]$ and $t' \in [0,t_1]$ are exchanged, then Eq.~(\ref{eqa10}) leads to the quantities
\begin{align}
&\quad \mathcal{I}^\pm_{{\textrm{Re}/\textrm{Im}}}(t_1,t_2,t_1+t_2) \nonumber\\
& \!=\!\left\{\!\!\!\!\begin{array}{rl}
&\!\!\frac{-1}{2\pi}\!\!\int_{-\infty}^{\infty}\!\! d\omega \,\sin (\omega t_2) {\textrm{Im}}[F(\omega, t_1) F'(-\omega, t_1)] S^+(\omega)\vspace{1ex}\\
&\!\!\frac{1}{2\pi}\!\!\int_{-\infty}^{\infty}\!\! d\omega \,\cos (\omega t_2) {\textrm{Re}}[F(\omega, t_1) F'(-\omega, t_1)] S^+(\omega)\vspace{1ex}\\
&\!\!\frac{\rm i}{2\pi}\!\!\int_{-\infty}^{\infty}\!\! d\omega \,\sin (\omega t_2) {\textrm{Re}}[F(\omega, t_1) F'(-\omega, t_1)] S^-(\omega)\vspace{1ex}\\
&\!\!\frac{\rm i}{2\pi}\!\!\int_{-\infty}^{\infty}\!\! d\omega \,\cos (\omega t_2) {\textrm{Im}}[F(\omega, t_1) F'(-\omega, t_1)] S^-(\omega)\vspace{1ex}
\end{array}\!\!.\right.\label{eqa11}
\end{align}

The key is to recognize -- specially  given the symmetry/anti-symmetry of $S^\pm (\omega)$ -- that for any given $t_1$,
\begin{align}
\nonumber {\textrm{Re}}[F(\omega, t_1) F'(-\omega, t_1)] S^\pm(\omega) & = \mathcal{F}[\mathcal{I}^\pm_{\textrm{Re}}(t_1,t_2,t_1+t_2)](\omega),\\
 \label{eqa45} {\textrm{Im}}[F(\omega, t_1) F'(-\omega, t_1)] S^\pm(\omega) & = \mathcal{F}[\mathcal{I}^\pm_{\textrm{Im}}(t_1,t_2,t_1+t_2)](\omega),
\end{align}
i.e., one can experimentally access the (inverse) Fourier transform of $ S^+(\omega)$ modulated by the real/imaginary part of $F(\omega, t_1) F'(-\omega, t_1)$. From this point of view, a natural way to do spectroscopy is as follows: (i) choose certain filter functions $F(\omega, t_1),F'(\omega, t_1)$; (ii) for a series of different $t_2$ values, perform experiments and reconstruct the values of $\mathcal{I}^\pm(t_1,t_2,t_1+t_2)$, and perform an even/odd extension to $t_2<0$ according to the symmetry/antisymmetry of $\mathcal{I}^\pm(t_1,t_2,t_1+t_2)$; (iii) perform, e.g., Discrete Fourier Transform (DFT) or Discrete-Time Fourier Transform (DTFT) on the collected values of $\mathcal{I}^\pm(t_1,t_2,t_1+t_2)$ and invert by the real/imaginary part of $F(\omega, t_1) F'(-\omega, t_1)$ to reconstruct the spectrum curve in certain frequency range; (iv) repeat (i)-(iii) for different $F(\omega, t_1),F'(\omega, t_1)$ to reconstruct the spectrum in different frequency ranges. 

The details on how the above can be done and what are the limits of the approach depend heavily on the experimental constraints. 

First, note that $t_2 \geq \Delta$ can only be sampled at multiples of $\delta.$ This implies that the Fourier transform can provide reliable information up to a frequency $\omega^{\rm Fourier}_{\rm co} = 2 \pi/ \delta.$ On the other hand, since in theory $t_2$ is not upper bounded (in the limit of purely dephasing noise as we consider here) one can in principle sample the spectrum with very high frequency resolution. We notice that in practice, when there is also non-dephasing noise, $t_2$ will be limited. Moreover, any new {\it time-trace}, i.e., value of $t_2$, implies a new experimental setup and thus adds to the overall cost of the characterization.   

Second, one notices that the reconstruction of $S^\pm(\omega)$ is ``windowed'' in a frequency domain by $F(\omega, t_1) F'(-\omega,t_1),$ and control constraints influence what sort of filter window can be generated. An obvious implication of the above is that $S^\pm(\omega)$ can only be inferred in the region where  $F(\omega, t_1) F'(-\omega,t_1)$ is non-zero. {To formalize this, we define the \textit{main frequency support} (MFS) of the chosen filters as the region $ \Omega: 0\leq\omega_a\leq \omega\leq \omega_b$ where 
\begin{equation*}
\begin{aligned}
&\int_{
\omega_a}^{\omega_b}
d\omega \Big|\mathcal{X}[F(\omega, t_1) F'(-\omega,t_1)]\Big| \\ \geq & \alpha\int_{0}^\infty
d\omega \Big|\mathcal{X}[F(\omega, t_1) F'(-\omega,t_1)]\Big|
\end{aligned}
\end{equation*}
and 
\begin{equation*}
\begin{aligned}
&\Big|\mathcal{X}[F(\omega, t_1) F'(-\omega,t_1)]\Big|_{\omega\in[\omega_a,\omega_b]} \\
\geq & \beta \max_{\bar\omega\in\mathbb{R}}   \Big|\mathcal{X}[F(\bar\omega, t_1) F'(-\bar\omega,t_1)]\Big|    
\end{aligned}
\end{equation*}
with $\mathcal{X} \in \{\textrm{Re}, \textrm{Im} \}$ and the factors $\alpha, \beta$ being some convenient values, e.g., $\alpha =1/2=\beta.$} Thus, to infer the value of $S(\omega)$ in a subregion of interest $\Omega$, one would like to generate a filter whose MFS is $\Omega.$ One can then infer the full $S^\pm (\omega)$ by sweeping $\Omega$ over the whole relevant frequency range. A second implication, is that one would like the filter to be sufficiently smooth in the region of interest, so as to facilitate the Fourier transform step, i.e., one does not want that the filter adds more frequency components to $S(\omega).$ The ideal scenario is then where $|F(\omega,t_1)|^2$ is a square window function of variable centre and width.

\begin{figure}
\centering
\includegraphics[scale=0.45]{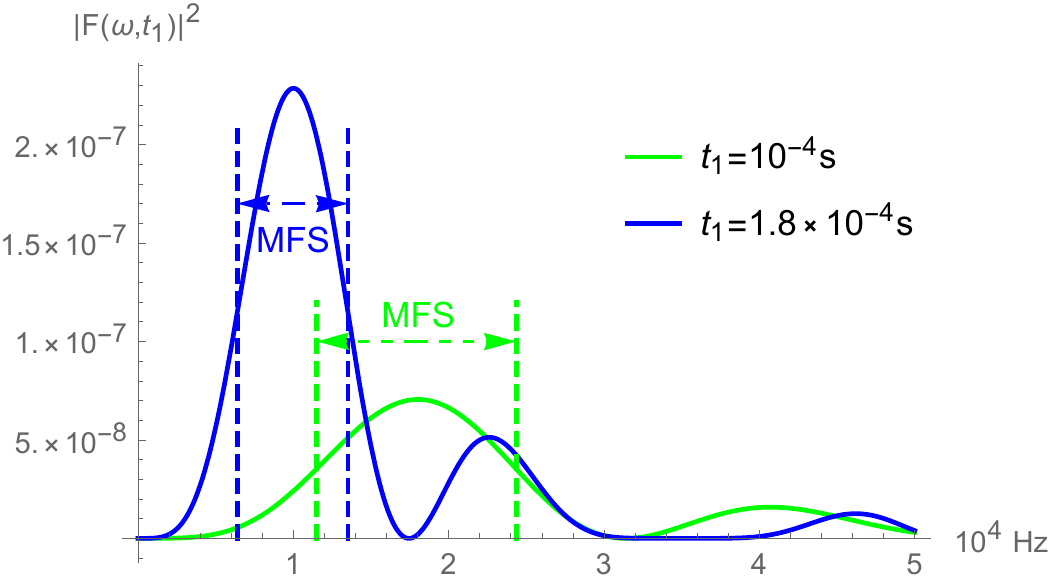}
\caption{Two examples of $1$-CPMG filters for $t_1=10^{-4}$s (green solid curve) and $t_1=1.8\times 10^{-4}$s (blue solid curve), with their corresponding MFSs marked (dashed). If these two filters are employed separately, the spectrum curve in the region $[0.64,2.43]\times 10^4$Hz can be recovered ultimately.}
\label{fig:cpmg}
\end{figure}

We will use DD sequences~\cite{hahn1950, ViolaDD, Khodjasteh2005,Khodjasteh2007,pazsilva2016} composed of $\pi$-pulse sequences (free evolution is the case where no $\pi$-pulses are applied) to generate $F(\omega,t_1)$ and $F'(\omega,t_1)$, since the filter function structure of such sequences is well understood and are thus practical if one is interested in setting their MFS in a controllable frequency range.  Notice that -- in the absence of knowledge about the noise -- a sufficiently powerful DD sequence can in principle mitigate the effect of the noise, i.e., ensuring that $A^+_{\vec{a}}$ is not close to zero, thus allowing us to extract information from observables. For example, one can use $n$-CPMG sequences made in the $[0,t_1]$ and $[t_2, T=t_2+t_1]$ intervals, implemented in a length-$t_1$ interval by applying pulses at $\{\tau = \frac{t_1}{2 (n+1)}, 3 \tau , \cdots , (2 n +1) \tau\}$ plus a pulse at the final time if $n$ is even. The key property is that by changing $t_1$, one can move the MFS of the filter function as desired, with only a minimal loss in power in the MFS (see {Fig}.~\ref{fig:cpmg}). Although it should be noted that the minimum switching time $\Delta$ implies that not every sequence can be used in the windows of length $t_1$. For example, for $t_1 < 2 \Delta$ only free-evolution can be considered in the window. Moreover, note that the minimum switching time imposes an upper bound to the frequency that can be uniquely sampled, namely $2 \pi/\Delta.$ Thus, to bypass this issue and avoid aliasing in our estimation, we will assume that the spectra under consideration have a frequency cut-off $\omega_{\rm co} \leq 2\pi/ \Delta.$ 

As information about the noise is gained, the choice of filter can be adapted to maximize the information gain in a given frequency band.  Moreover, note that DD sequences provide a systematic way to produce filters with varying MFSs, that can eventually be glued together as needed. This can be taken further, however. Indeed, one can build filters with optimal spectral concentration in a window of interest using $\pi$-pulse sequences, as was shown in \cite{frey2020}, which shares the key property of Slepian functions, i.e., being concentrated in both frequency and time.

Technical details of performing this Fourier transform method are presented in Appendix \ref{app:para}.

\subsection{Full-access reconstruction}

One can take the above further. Recall, we have in principle access to the values of the integrals $ \mathcal{I}^\pm_0
(0, \delta,q \delta , (q + 1) \delta)$, for $q\geq \lceil \Delta/\delta\rceil$. Provided that $\delta$ is sufficiently small and that $C^\pm(t,t')$ is sufficiently smooth so that it can be taken approximately constant in the small integration domain, one can then assume that this implies access to the piecewise constant $C^\pm( \frac{\Delta+ s  \delta}{\sqrt{2}},0)$, for $0\leq s\in \mathbb{N}$.  With this, one can then use classical processing to build
$$M_f =  \sum_{s} f(s) C^\pm( \frac{\Delta+ s  \delta}{\sqrt{2}},0) $$
for a suitable $f(s),$ so that $M_f$ provides information about a feature of interest. For example, choosing $f(s) = e^{-{\rm i} W s \delta}$ and having data for sufficiently large values of $s$, would yield $\mathcal{F} [C^\pm( \frac{\Delta+ s \delta}{\sqrt{2}} ,0) ] (W),$ i.e., the value of the spectrum at $\omega =W.$ The freedom in $f(s)$ and the seemingly unlimited ability to exploit $M_f$ is nothing but a reflection of the great level of detail control can provide. 

This level of detail in the time domain, however, comes at a cost. Inferring the value of $ \mathcal{I}^\pm_0
(0, \delta,q \delta , (q + 1) \delta)$ requires nonlinear combinations of expectation values resulting from a number of control setups (choice of observable, control sequence, $t_2$, and $t_1$) that scales with $1/\delta.$ Hence, while one gets more detail, the error in the estimation of each integral also grows. We present it here for completeness since it may be of use if the interest is to access a particularly fine feature in the bath correlation functions, but leaving for future work going over the practicalities. As for the specific algorithm used to reconstruct the spectrum curve from values of $ \mathcal{I}^\pm_0
(0, \delta,q \delta , (q + 1) \delta)$ or more generally from observable expectation values, there are certainly multiple choices that can be developed. For example, Bayesian estimation algorithms \cite{ferrie2018} are particularly helpful when the measurement shot number is not large and a prior structure of the spectrum function is given. In Sec. \ref{sec:simu}, we focus on simulating our Fourier-based method that requires no prior knowledge on the structure of the spectrum.

\subsection{Overcoming the multivalue problem}\label{subsec:relation}

An additional benefit of the frequency domain, is that it provides the means to overcome the multivalue problem. Concretely, it allows one to establish a bound relating the $c$ spectrum and its $q$ counterpart, which can then be leveraged for noise estimation purposes. 

\subsubsection{Relationship between $c$ and $q$ spectra}

A superficial deduction would suggest that the $c$ and $q$ components of the noise statistics are only loosely related. Two observations support this. First, there are scenarios where the $c$ spectrum can be any physically-allowed function while the $q$-noise is zero. This is the case, for example, when the noise is classical, i.e., when $B(t)=b(t)B$ with $b(t)$ a $c$-stochastic process and $B$ an arbitrary non-zero time-independent bath operator. Second, if one considers an environment consisting of a single qubit such that $B(t) = b_x(t) \sigma_x + b_y(t) \sigma_y,$ then the relations $[\sigma_x,\sigma_y]_-=2{\rm i}\sigma_z$ and $[\sigma_x,\sigma_y]_+=0$ indicate that it is possible for the $q$-correlation function $\langle [B(t),B(t')]_-\rangle \neq 0$ at $(t,t')=(t_a,t_b)$ while the $c$-correlation $\langle [B(t),B(t')]_+\rangle = 0$ at $(t_a,t_b)$, i.e., the opposite behaviour as in the first observation. Thus, establishing a clear relationship in the time domain does not seem straightforward. 

In the frequency domain, however, one can prove the following result.

\begin{theorem}\label{theorem2}
Let the bath operator $B(t)$ and bath state $\rho_B$ acting on a Hilbert space $\mathcal{H}$ describe a wide-sense stationary noise process. Let any operator on $\mathcal{H}$ then be represented by $X = \int_{P,Q} dp dq X_{pq}
\ket{p} \bra{q}.$ If at least one of the following is satisfied, 
\begin{align}
&\int_{\mathbb{R}}dt\int_P dp\Big|\langle   \bra{q} B(t) \ket{p} \!\bra{p} B(0) \ket{q} 
\rangle_c\Big|<\infty,\notag\\
&\int_P dp\int_{\mathbb{R}}dt\Big|\langle \bra{q} B(t) \ket{p} \!\bra{p} B(0) \ket{q}\rangle_c\Big|<\infty, \label{eq39}\\
&\int_{\mathbb{R}\times {P}}d(t,p)\Big|\langle \bra{q} B(t) \ket{p} \!\bra{p} B(0) \ket{q}\rangle_c\Big|<\infty\notag
\end{align}
for arbitrary $|q\rangle\in\mathcal{H}$ and resolution of identity $I=\int_P dp|p\rangle\langle p|$ on $\mathcal{H}$, and provided $S^\pm(\omega)\equiv\mathcal{F}\big[\braket{[B(t),B(0)]_\pm}\big]$ exist, then
\begin{equation}\label{eq41}
\Big|S^-(\omega)\Big|\leq S^+(\omega)
\end{equation}
for all $\omega\in\mathbb{R}$. 

Two useful corollaries follow from Eq. \eqref{eq41}. Namely, we further have
\begin{equation}\label{eq40}
\Big|\langle[B(t),B(t')]_-\rangle\Big|\leq \langle[B(0),B(0)]_+\rangle
\end{equation}
for all $t,t'\in\mathbb{R}$, and
\begin{align}
&\quad 2\Big|\int_{T_1}^{T_1+T_2}dt\int_0^{T_1}dt' \langle[B(t),B(t')]_-\rangle\Big|\notag\\
&\leq \!\Big(\!\int_0^{T_1}\!dt\!\int_0^{T_1}\!dt'\!+\!\int_0^{T_2}\!dt\!\int_0^{T_2}\!dt'\Big)\langle[B(t),B(t')]_+\rangle\label{eq42}
\end{align}
for arbitrary $T_1,T_2\geq 0$. 

Further, it follows that given arbitrary real antisymmetric $\hat{S}^-(\omega)$ and symmetric $\hat{S}^+(\omega)$ such that $|\hat{S}^-(\omega)|\leq \hat S^+(\omega)$ holds for all $\omega \in\mathbb{R}$, then, assuming that $\mathcal{F}^{-1}\big[\hat{S}^\pm (\omega)\big]$ exist, there exist $\hat{\mathcal{H}}$, $\hat\rho_B,$ and $\hat B(t)$ capable of generating a wide-sense stationary process with such spectra $\hat{S}^\pm(\omega)$.
\end{theorem}

While full details of the proof are provided in the Appendix \ref{app:theorem}, we highlight that the conditions in Eq. \eqref{eq39} require that the bath correlation functions decay sufficiently fast, so that one can employ the Fubini-Tonelli theorem in a key step of the proof. 
Moreover, Theorem \ref{theorem2} means a sufficient and necessary condition for QNS results to be physical is Eq. \eqref{eq41} under mild assumptions, 
which establishes a point-wise hierarchy relating the $c$ and $q$ spectra. Thus, after performing QNS and obtaining estimates for $\hat{S}^\pm(\omega)$, one needs to check Eq. \eqref{eq41} to ensure that it holds for all $\omega\in\mathbb{R}$ and that one has physically consistent estimation. 

The existence of Eq. \eqref{eq41} is not without any hint in history. For example, in thermal equilibrium, the fluctuation-dissipation theorem shows $S^+(\omega)=\coth({\hbar \omega/2k_BT})S^-(\omega)$ \cite{callen1951,kubo1957}. Using Fermi's golden rule, Ref. \cite{schoelkopf2003,quintana2017} establish a relationship at zero tilt: $S^-(\omega)=(1-2p_{\rm stray})S^+(\omega)$ where $p_{\rm stray}$ is the steady-state population. These relationships are clearly consistent with Eq. \eqref{eq41}, and so are the experimentally reconstructed spectra in Ref. \cite{shen2023}.

As should be expected, the two corollaries are weaker statements than Eq. \eqref{eq41}. For example, Eq. \eqref{eq40} is insufficient for Eq. \eqref{eq41}. To see this imagine two independent bath operators $B(t),\tilde{B}(t)$, with $\rho_B=\rho_{\tilde{B}}=|1\rangle\langle 1|$. Let $B(t)=x_{12}(t)(|1\rangle\langle 2|+|2\rangle\langle 1|)+y_{12}(t)(-{\rm{i}}|1\rangle\langle 2|+{\rm{i}}|2\rangle\langle 1|)$ with $\langle x_{12}(t)y_{12}(t')\rangle_c=\sin[\omega_a(t-t')]$, and $\tilde{B}(t)=\tilde{z}_{1}(t)|1\rangle\langle 1|$ with $\langle \tilde{z}_1(t)\tilde{z}_1(t')\rangle_c=2\cos[\omega_b (t-t')]$. Then one can verify that $|\langle[B(t),B(t')]_-\rangle|\leq \langle[\tilde{B}(0),\tilde{B}(0)]_+\rangle$. However, $S^-(\omega)=4\pi[\delta(\omega-\omega_a)-\delta(\omega+\omega_a)]$ and $\tilde{S}^+(\omega)=4\pi[\delta(\omega-\omega_b)+\delta(\omega+\omega_b)]$. Hence, when $|\omega_a|\neq|\omega_b|$, we have $|S^-(\omega_a)|>\tilde{S}^+(\omega_a)$, invalidating Eq. \eqref{eq41}.
On the other hand, Eq. \eqref{eq42} is even weaker, which can be seen by taking one of $T_1,T_2$ to be zero. Nevertheless, this inequality might still be useful in practical experiments, as its 2D time integral form often appears explicitly in the expression of $E[ \hat O(T)]$, and is thus useful as a cross check for inconsistencies in the measurement data.

In addition to providing a consistency-check mechanism for QNS results, Theorem \ref{theorem2} is fundamental to overcoming the multi-value problem, as we now demonstrate.

\subsubsection{A solution to the multivalue problem}\label{subsec:multi-value}

Let us start by summarizing the problem. The $q$ component of the spectrum influences the dynamics of a quantum system via integrals $\mathcal{I}^-$ which are in turn the arguments of periodic functions. This implies that -- simplifying considerations absent -- the value of any such integral can only be inferred up to a factor of $2 \pi.$ This clearly limits our ability to estimate the $q$ component of the noise in the strong coupling limit, when the values of typical $\mathcal{I}^-$ integrals may exceed $2 \pi.$ As detailed in Sec.~\ref{subsub:multi}, the problem reduces to  estimating $\mathcal{I}^-(\vec{T})$, given $s(t_2) \equiv \sin({\rm Im}[\mathcal{I}^-(t_1,t_2,t_1+t_2)])$ and $c(t_2) \equiv \cos({\rm Im}[\mathcal{I}^-(t_1,t_2,t_1+t_2)])$, and fixed $F(\omega,t_1),F'(\omega,t_1).$ To uniquely determine $\mathcal{I}^-(\vec{T})$, one can apply the following strategy.

First, let us recall that the estimation of the $c$-spectrum can be done independently of the $q$-spectrum, in the sense that one can isolate $\mathcal{I}^+_{{\textrm{Re}/\textrm{Im}}}(\vec{T})$ regardless of the multivalue problem. Thus, one can assume an accurate estimate of $S^+(\omega)$ is available, which we denote $\hat{S}^+(\omega).$ It follows then that
\begin{align*}
&\quad|\mathcal{I}^-(t_1,t_2,t_1+t_2)|\!\\
& \leq  \frac{1}{\pi} \!\int_{0}^{\infty}\!\!\! d\omega | e^{{\rm i} \omega t_2 }F(\omega, t_1)F'(-\omega, t_1)| |S^-(\omega)|\\
& \leq \frac{1}{\pi}  \! \int_{0}^{\infty}\!\! \! d\omega | e^{{\rm i} \omega t_2 }F(\omega, t_1)F'(-\omega, t_1)| S^+(\omega)\\
&\equiv \bar{\mathcal{I}}^+(t_1,t_2,t_1+t_2),
\end{align*}
One can then define a {\it safe zone} in which the multi-value problem is non-existent. That is, a range of values for $t_2$ where, for fixed $e^{{\rm i} \omega t_2} F(\omega, t_1)F'(-\omega, t_1)$, knowledge of $s(t_2)$ and $c(t_2)$ provides a unique estimate of $\mathcal{I}^-(t_1,t_2,t_1+t_2).$  The trivial solution is when $|{\rm Im}[\mathcal{I}^-(t_1,t_2,t_1+t_2)]| \leq \pi/2,$ but can also include domains in which the range of ${\rm Im}[\mathcal{I}^-(t_1,t_2,t_1+t_2)]$ is in a known iteration of the $[-\pi/2,\pi/2]$ interval, the fundamental domain of the $\arcsin$ function. 

The trivial safe zone can be found as follows. Given knowledge of $S^+(\omega),$ one can numerically search for an interval $[t_a,t_b]$ such that $\bar{\mathcal{I}}^+(t_1,t_2,t_1+t_2) < \pi/2$ when $t_2\in [t_a,t_b]$. For a general $S^+(\omega),$ this can happen for ``small" $t_2,$ i.e., when $\Omega_{\rm cutoff} t_2 \ll 1$, with $\Omega_{\rm cutoff}$ the effective cut-off frequency of $| e^{{\rm i} \omega t_2} F(\omega, t_1)F'(-\omega, t_1)| S^+(\omega),$ or ``large" $t_2,$ i.e., when $e^{{\rm i} \omega t_2}$ oscillates (as a function of $\omega$) much faster than $ F(\omega, t_1)F'(-\omega, t_1) S^+(\omega)$, but can be in an intermediate regime given a specific $S^+(\omega)$. However, while in the above scenarios one can safely estimate $\mathcal{I}^-(t_1,t_2,t_1+t_2)$ by using $\arcsin $ when $t_2\in[t_a,t_b]$, in practice one requires knowledge of $\mathcal{I}^-(t_1,t_2,t_1+t_2)$ for a wide -- ideally infinite -- range of $t_2$ values in order to be able to perform a Fourier transform and obtain an estimate for \eqref{eqa45}.  Thus, one would like a mechanism in which the trivial safe zone $[t_a,t_b]$ can be systematically extended. 

\begin{figure}
\centering
\includegraphics[scale=0.341]{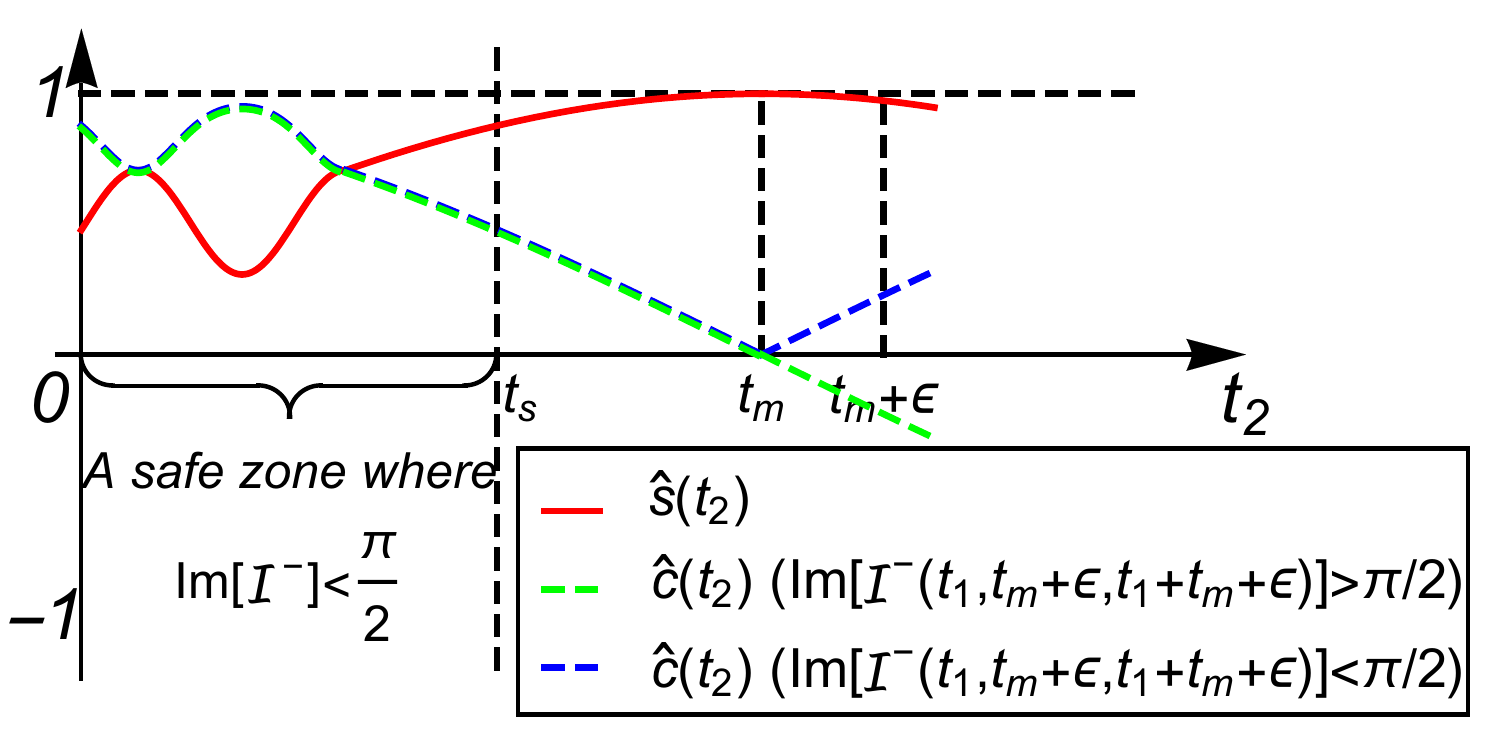}
\caption{An example to solve the multi-value problem. Here $[0,t_s]$ is an original safe zone where ${\rm Im}[\mathcal{I}^-(t_1,t_2,t_1+t_2)]<\pi/2$, and thus $\arcsin$ can be directly performed on $\hat{s}(t_2)$. When $t_2$ exceeds $t_m$, the task is to judge whether ${\rm Im}[\mathcal{I}^-(t_1,t_2,t_1+t_2)]$ exceeds ${\pi}/{2}$ or not, which can be determined by the sign of the value of $\hat{c} (t_m+\epsilon)$.}
\label{f01}
\end{figure}

To do so, imagine now one has identified one of these values, say  $t_2=t_s$, in the trivial safe zone. Then, one samples $t_2$ in increments of some $\epsilon$ until the estimate $\hat{s}(t_2)$ approaches $\pm 1,$ say at $t_2 = t_m.$ Take Fig.~\ref{f01} for example, where $\hat{s}(t_2=t_m)=1$ and thus $[t_a,t_b]=[0,t_m]$. At this point, assuming the sampling step-size $\epsilon$ is sufficiently small to ensure that $\mathcal{I}^-(t_1,t_2,t_1+t_2)$ has been ``continuously" sampled as a function of $t_2,$ one can assess  the range of ${\rm Im}[\mathcal{I}^-(t_1,t_2,t_1+t_2)]$ when leaving the original safe zone by monitoring the experimentally inferred value of $c(t_2) = \cos({\rm Im}[\mathcal{I}^-(t_1,t_2,t_1+t_2)])$. If $\hat{c}(t_m-\epsilon) \hat{c}(t_m+\epsilon) >0,$ it must be that ${\rm Im}[\mathcal{I}^-(t_1,t_m+\epsilon,t_1+t_m+\epsilon) ]< \pi/2,$ i.e., one realizes that $\hat{\mathcal{I}}^-(t_1,t_m+\epsilon,t_1+t_m+\epsilon)={\rm i}\arcsin(\hat{s}(t_m+\epsilon))$. In contrast, if $\hat{c}(t_m-\epsilon) \hat{c}(t_m+\epsilon) < 0,$ it must be that ${\rm Im}[\mathcal{I}^-(t_1,t_m+\epsilon,t_1+t_m+\epsilon)] > \pi/2$, i.e., one has that $\hat{\mathcal{I}}^-(t_1,t_m+\epsilon,t_1+t_m+\epsilon)={\rm i}[\pi-\arcsin(\hat{s}(t_m+\epsilon))]$. In both cases, one knows how to correctly determine $\mathcal{I}^-(t_1,t_2,t_1+t_2)$, and thus has expanded the original safe zone from $[0,t_m]$ to $[0,t_m+\epsilon]$. One can then continue the sampling of $t_2$, and decide at the next point where $\hat{s}(t_2 = t'_m) =\pm 1$ if the value of $\mathcal{I}^-(t_1,t_2,t_1+t_2)$ has moved to the next iteration of the fundamental domain by assessing the negative/positive character of $\hat{c}(t_m' -\epsilon) \hat{c}(t_m' +\epsilon)$. In this way, one can systematically extend the safe zone and keep track of in which iteration of the fundamental domain $\mathcal{I}^-(t_1,t_2,t_1+t_2)$ is for a sampled $t_2.$ As such,  $ \hat{\mathcal{I}}^-(t_1,t_2,t_1+t_2)={\rm i}[k\pi+(-1)^k\arcsin (\hat{s}(t_2))]$ when $\mathcal{I}^-(t_1,t_2,t_1+t_2)$ is in the $k$-th iteration of the fundamental domain provides a unique estimate, as desired.

\subsection{Limits and comparisons with other methods}

Let us now discuss what are the limits of our method. Since we use Fourier analysis by sampling $t_2$ via Eq.~\eqref{eqa45}, Nyquist-Shannon sampling theorem implies that our method can resolve a frequency range up to ${1}/{2 \Delta}$ (Hz) where $\Delta$ is the minimum pulse separation time implemented. We highlight that this is true even in the situation were a minimal time for the first pulse, say $\tau_d>\Delta$ exists. Notice that this apparent obstacle can be overcome by using stationarity to move $\tiny{\begin{pmatrix}
\cdot & \cdot & 0\\
\cdot & 0 & 0\\
0 & 1 & 0
\end{pmatrix}}^\pm$ to a proper 
$\tiny{\begin{pmatrix}
\cdot & \cdot & 0\\
\cdot & 0 & 1\\
0 & 0 & 0
\end{pmatrix}}^\pm$
in the language of Sec. \ref{cla}.

On the other hand, the lower frequency limit of our method is $1/\max(t_2)$, which can be easily seen from a discrete Fourier transform point of view on Eq. \eqref{eqa45}. Interestingly, in the strictly dephasing scenario we consider here, despite the existence of a $\mathbb{T}_2^\alpha$ time, i.e., the time $t_2=\mathbb{T}_2^\alpha$ for which $A^+_{(1,1,1)} = 1/2$ for a given pulse sequence $\alpha$, our method in principle allows an unrestricted reconstruction of the $c$ spectrum. This can be seen from noting that the combinations of $Q_1^\pm, Q_5^\pm$ and $Q_6^\pm$ quantities (see Sec. \ref{cla}) give us direct access to $\tiny{\begin{pmatrix}
\cdot & \cdot & 0\\
\cdot & 0 & 0\\
0 & 0 & 1
\end{pmatrix}}^\pm$ and these signals are only ever suppressed by $A^+_{(s_1,0,s_2)}$, i.e., the exponents only involve integrals over a domain of size $t_1$ rather than $t_2$.  In practice, when the noise is not purely dephasing, $t_2$ is effectively bounded by the energy relaxation time $\mathbb{T}_1$, which sets the lower frequency limit as about $1/\mathbb{T}_1$. In contrast, the situation with the $q$ spectrum is more delicate, since quantities containing $\tiny{\begin{pmatrix}
\cdot & \cdot & 0\\
\cdot & 0 & 0\\
0 & 1 & 0
\end{pmatrix}}^-$ are suppressed by $A^+_{(s_1,1,s_2)},$ which contain integrals explicitly depending on $t_2$ and can be suppressed by the exponential factor for large $t_2.$ Thus, it would seem that in the $q$ component case $t_2$ is effectively bounded. Notice, however, that stationarity implies that $\tiny{\begin{pmatrix}
\cdot & \cdot & 0\\
\cdot & 0 & 0\\
0 & 1 & 0
\end{pmatrix}}^- =\tiny{\begin{pmatrix}
\cdot & \cdot & 0\\
\cdot & 0 & 1\\
0 & 0 & 0
\end{pmatrix}}^-,$ and the latter can be recovered via $Q_1^\pm, Q_5^\pm, Q_6^\pm$ and thus, as exemplified in Sec. \ref{subsec:equs}, it implies that estimation of the $q$ component is also {\it not} $\mathbb{T}_2^\alpha$-limited. 

We are now in a position to make concrete comparisons with the existing methods in literature. First, standard spectroscopy experiments with $\pi$-pulse control sequences -- e.g., Hahn echo and CPMG sequences -- reveal the $c$-spectrum curve roughly above $1/\mathbb{T}_2^\alpha$ \cite{yuge2011,muhonen2014,chan2018,yoneda2018}. There are various methods aiming at overcoming this limit, \cite{bialczak2007,lanting2009,sank2012}. A notable one is to periodically repeat Ramsey experiment (allowed to be decorated with DD sequences) and single-shot measurement on the probe qubit, perform Fourier transform on the binary time series record and average the reconstructed spectrum curves obtained from different time series \cite{yan2012}. The upper cutoff frequency is $\frac{1}{2 dt}$ where $dt$ is the consecutive measurement time difference. The lower cutoff frequency is the reciprocal of the total acquisition time which can exceed $\mathbb{T}_1$. Hence their method can enter a lower frequency region of the $c$-spectrum even in systems with a short $\mathbb{T}_1$. { On the flip side, since $dt$ is often significantly larger than $\Delta$ \cite{yan2012,yoneda2018}, this implies that the high frequency cut-off of this method is considerably smaller. For example, there is a frequency gap of about four orders of magnitude between this Ramsey measurement-based method and the CPMG method (the Comb method with a single frequency comb) in \cite{yoneda2018}.}
We note that the repeated Ramsey measurement method in \cite{yan2012} was further analyzed in \cite{wudarski2022} to obtain Eq. (26) therein, the same as Eq. \eqref{eqa10}'s $c$-spectrum part in our paper. Regretfully, \cite{wudarski2022} did not develop a systematic approach based on this formula to reconstruct arbitrary unknown $c$-spectrum, but only developed the formulas for several specific spectra types. Furthermore, a common feature in the above existing works is that the role and effect of $q$-noise is absent.

\subsubsection{Comparison with existing techniques in NMR employing non-$\pi$ pulses}

The application of non-$\pi$ pulses in NMR has a long history and significant success in aiding to resolve material structure. For example, Stimulated Echo (STE) \cite{mims1972} has the same pulse sequence as our low-frequency protocol (e.g., Eqs. \eqref{eqx11} and \eqref{eqa1}), with three 90 degree pulses (including the state preparation pulse $|\pm\rangle_z \rightarrow |\pm\rangle_y$ that is omitted in this paper) followed by a projective measurement, generating the same sandwich structure of a middle interval ($[t_1,t_2]$ in our language), where the switching function takes zero value and the spin captures no phase \cite{klauder1962}, surrounded by two phase-capturing intervals ($[0,t_1]$ and $[t_2,T]$). As a result, STE and the developed other techniques such as HYSCORE \cite{hofer1986} have wide applications in volume-selective spectroscopy, diffusion spectroscopy, MRI, etc. In contrast, here we extend the system model from the semi-classical one to a full quantum-mechanical version \eqref{basicH} and reestablish the technique, reminiscent of the formulation of DD in quantum information on the basis of existing pulse techniques in NMR. Therefore, the wide-range classical and quantum-mechanical spectra of the bath sensed by a single qubit can now both be reconstructed after subtracting the influence of each other, and the system dynamics can be more accurately characterized and predicted when richer control settings (beyond the dephasing-preserving scenario) are considered.

\section{Simulation}\label{sec:simu}
We demonstrate our protocol with detailed numerical simulations, where we explore - among other things - the effect of the error induced by a limited number of measurements. Concretely, we simulate the measurement outcomes resulting from a bath with $1/f$ spectra decorated with some ``bumps'' as 
\begin{equation}
S^\pm(\omega)|_{\omega > 0}=\frac{a^\pm_1}{1+a^\pm_2\omega}+ \sum_{j=1}^{N^\pm}\frac{b_j^\pm}{1+c_j^\pm(\omega-\omega_j^\pm)^2}, 
\end{equation}
and $S^\pm (\omega)|_{\omega < 0}=\pm S^\pm(-\omega).$ The parameters are tuned such that Eq. \eqref{eq41} is satisfied and the $c$-spectrum resembles the curve in the experimental result in Ref. \cite{chan2018}. Specifically they are $a_1^+/\hbar^2=240\ {\rm kHz}$, $a_2^+=0.02\ {\rm s}^2$, $N^\pm=3$, $\bar b_1^+=7\ {\rm kHz}$ ($\bar b_i^\pm\equiv b_i^\pm/\hbar^2$), $\bar b_2^+=8\ {\rm kHz}$, $\bar b_3^+=3\ {\rm kHz}$, $c_1^+=0.002\ {\rm s}^2$, $c_2^+=200\ {\rm ms}^2$, $c_3^+=33.33\ {\rm ms}^2$, $f_1^+=400\ {\rm Hz}$ ($f_i^\pm\equiv\omega_i^\pm/(2\pi)$), $f_2^+=1\ {\rm kHz}$, $f_3^+=3.5\ {\rm kHz}$. $a_1^-/\hbar^2=125\ {\rm kHz}$, $a_2^-=0.0125\ {\rm s}^2$, $\bar b_1^-=8\ {\rm kHz}$, $\bar b_2^-=650\ {\rm Hz}$, $\bar b_3^-=600\ {\rm Hz}$, $c_1^-=0.04\ {\rm s}^2$, $c_2^-=156.25\ {\rm ms}^2$, $c_3^-=123.46\ {\rm ms}^2$, $f_1^-=390\ {\rm Hz}$, $f_2^-=1.6\ {\rm kHz}$, $f_3^-=3\ {\rm kHz}$. Furthermore, we add a constant value $S_{\rm white} =(2\pi\omega)^{1/4}-70\sqrt{2}\pi^{1/4}$ to $S^+(\omega)$ in the region $\omega/(2\pi)\geq 20\ {\rm kHz}$ to simulate the white noise floor in \cite{chan2018}. Finally, in the region $\omega\geq 50\ {\rm kHz}$, we multiply $S^-(\omega)$ with $\cos(\pi+ 3.95\times 10^{-4}\omega)$ to add some fluctuation features to the curve. We set the high frequency cutoff of the spectra as $60\ {\rm kHz}$.

\begin{figure*}
\centering
\includegraphics[width=\textwidth]{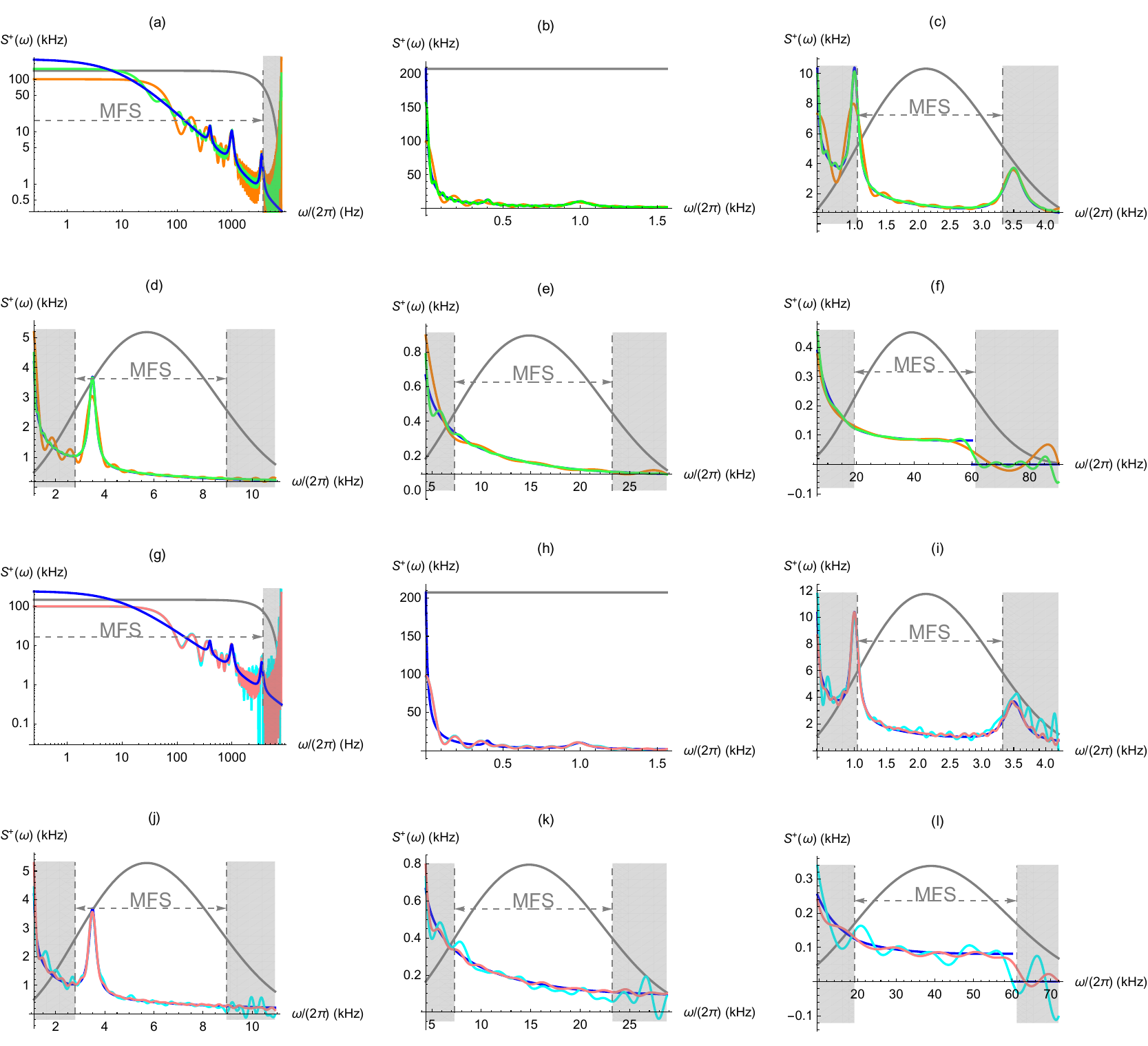}
\caption{Simulated QNS result for a $c$-spectrum (blue curve). Subplots (a)-(f) and (g)-(l) explore the effect of finite $K$, the number of time traces (with $t_2=kT_s|_{k=0,1,...,K}$), and the effect of the number of shots, respectively. The relevant frequency region given by the MFS of the chosen filter (gray curve for $|F(\omega,t_1)|^2$ re-scaled to fit in the plot) is highlighted in each subplot. Subplot (a)((g)) is the log-plot of (b)((h)), with a filter generated by $t_1=0.12\ {\rm ms}$ period of free evolution.  On the other hand, (c)-(f) employ Hahn echo sequences of length $t_1=350,130,50,19\ {\rm \mu s}$, respectively. The sampling periods for (b)-(f) are $T_s=(2/3,4/15,4/15,1/5,1/5)t_1$, respectively. $K$'s for the green plots in (b)-(f) are 
$250,80,120,50,30$, and a third of those for the orange curves. $T_s$ and $K$'s in (g)-(l) are the same as the green curves but in each case with a different number of shots per measurement. Concretely, pink (aqua) curves cost $10^6,10^5,10^6,10^7,10^7$ ($10^4,10^4,10^5,10^6,10^6$) measurement shots for each observable in the five frequency regions respectively.}
\label{simu1}
\end{figure*}

We start by employing our Fourier Transform Method (with the first and third intervals sharing the same $\pi$-pulse sequences for simplicity, i.e., $Y^-_{(1,0,0)}(t)=Y^-_{(0,0,1)}(t+t_2)$) to reconstruct the $c$-spectrum (divided by $\hbar^2$). The results are shown in Fig. \ref{simu1}. Subplots (a)-(f) explore the effect of a finite number of time traces (values of $t_2$) in the different frequency regions associated to the sequences used and their corresponding MFS, but assume an infinite number of shots per measurement. The more resource intensive implementation (green) requires 530 different experimental setups (combinations of DD sequences and $t_2$) and the spectrum estimation shows excellent agreement with the simulated noise. {While cutting the number of setups in principle leads to a loss in performance (orange), cutting them by a factor of three did not have a significant impact in our simulations.} On the other hand, subplots (g)-(l) explore the effect of a finite number of shots per measurement, i.e., the finite sampling error. The effect of finite measurements is more appreciable (in terms of the absolute value of the error) in the very {low} frequency range ($\omega < 1$ kHz), in the limit of applicability of the $Y^-_{(1,0,0)}(t)=Y^-_{(0,0,1)}(t+t_2)$ of our protocol given the $\delta,\Delta$ constraints. This can be improved in principle by optimizing the $Y^-_{(1,0,0)}(t)\neq Y^-_{(0,0,1)}(t+t_2)$ scenario. 
\begin{figure*}
\centering
\includegraphics[width=\textwidth]{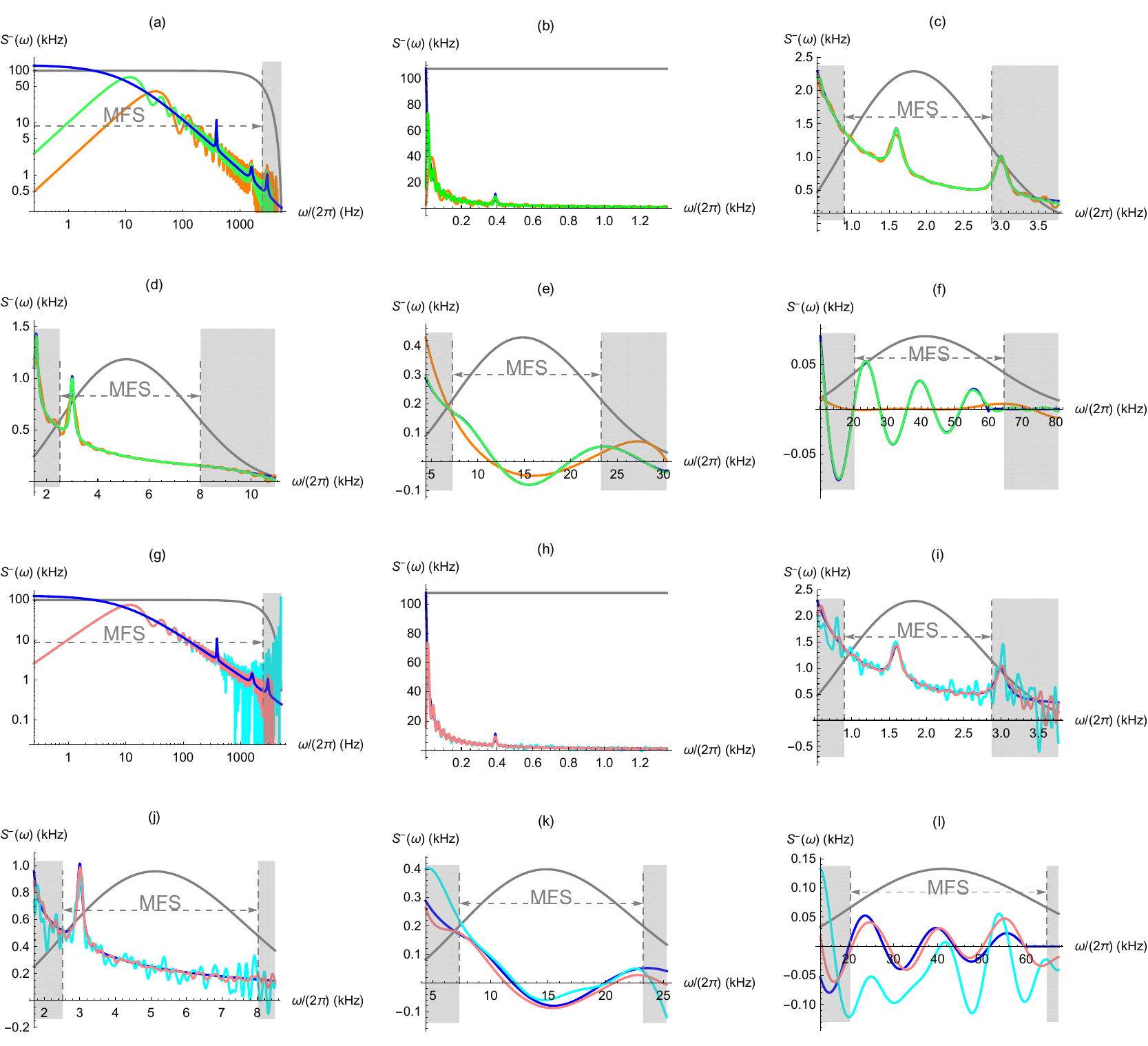}
\caption{Simulated QNS result for a $q$-spectrum (blue curve). Subplots (a)-(f) and (g)-(l) explore the effect of finite $K$, the number of time traces (with $t_2=kT_s|_{k=0,1,...,K}$), and the effect of the number of shots, respectively. The relevant frequency region given by the MFS of the chosen filter (gray curve for $|F(\omega,t_1)|^2$ re-scaled to fit in the plot) is highlighted in each subplot. Subplot (a)((g)) is the log-plot of (b)((h)), with a filter generated by $t_1=0.18\ {\rm ms}$ period of free evolution.  On the other hand, (c)-(f) employ Hahn echo sequences of length $t_1=405,145,50,18\ {\rm \mu s}$, respectively. The sampling periods for (b)-(f) are $T_s=(4/5,1/5,1/5,1/5,1/5)t_1$, respectively. $K$'s for the green plots in (b)-(f) are 
$240,140,200,18,30$, and a third of those for the orange curves. $T_s$ and $K$'s in (g)-(l) are the same as the green curves but in each case with a different number of shots per measurement. Concretely, pink (aqua) curves cost $10^6,10^6,10^7,10^7,10^8$ ($10^4,10^5,10^6,10^6,10^7$) measurement shots for each observable in the five frequency regions respectively.}
\label{simu2}
\end{figure*}

Once the $c$-spectrum is estimated, we reconstruct the $q$-spectrum. As seen from Fig.~\ref{simu2}, our method can also faithfully estimate it. The settings in both Fig. \ref{simu1} and \ref{simu2} satisfy the requirements \eqref{eqAppC3} and \eqref{eqf4}. The behaviour and performance for the $q$-spectrum is similar to the $c$-spectrum in most aspects, with some exceptions. Mainly in the very low frequency range (around $\omega=0$), we reach the limit of applicability of our protocol as in the $c$-spectrum estimation. In contrast to the $c$-estimation, however, $\omega=0$ is often a discontinuous point for a $1/f$ type $q$-spectrum while DTFT, which is adopted by our simulation, must output a continuous function on frequency. Further note that in the highest frequency range, the orange curve in (f) fails to capture the oscillations. This is mainly because the added $\cos(\pi+ 3.95\times 10^{-4}\omega)$ fluctuation in our model being highly concentrated in the frequency domain and easily missed by the modulation term $\sin(\omega t_2)$ generated from merely $10$ time traces therein. The estimation error is generally larger in (l) than in other subplots, due to weaker spectrum strength and thus higher relative sensitivity to noise and errors in measurement.

\section{Conclusions}\label{sec:con}
Employing a single-qubit probe system, we introduced a quantum noise spectroscopy protocol capable of faithfully reconstructing both the low-frequency and the non-classical aspect of a dephasing Gaussian bath spectra. Our protocol relies on a systematic investigation and employment of non-$\pi$ pulses on the qubit probe, which allows to generate zero-valued effective switching functions concentrating the long-time information of the bath correlation function, and endows us with an exact system evolution equation revealing the effect of non-classical bath. We exploit these equations to propose our exact noise characterization protocol and demonstrate its effectiveness through detailed simulations. 

Moreover, under mild assumptions,  we proved a sufficient and necessary characterization on the relationship between classical and non-classical spectra from the same bath, which helps to solve the multi-value problem to determine the non-classical noise, and also can be necessary and useful in curbing QNS results to be physical.

As a step towards a QNS protocol completely characterizing a general decoherence process from an arbitrary bath, our results reveal the crucial role of non-$\pi$ pulses in performing a broadband QNS task, and systematically extend the current single-qubit QNS working regime to the scenario where the non-classical nature of the bath needs to be considered or investigated. We expect this work to be a useful tool to fill in the low-frequency estimation gaps in current QNS results and to investigate the quantum nature of practical baths. It would also be valuable to extend our protocol to more general scenarios, such as baths with non-Gaussian features or scenarios including noise from the control signals.

\section*{Acknowledgements}  Work at Griffith was supported (partially) by the Australian Government through the Australian Research Council’s Discovery Projects funding scheme (Project No. DP210102291) and via AUSMURI Grant No. AUSMURI000002. Yuanlong Wang was also supported by the National Natural Science Foundation of China (No. 12288201).

\appendix

\section{Calculation of the cumulants of $V_{\hat o,\vec{r},{\vec{r}'}}(T)$}\label{app:cumulant}

We denote $\tilde{H}(t)\equiv y_{\vec{r}}(t) \sigma_z \otimes B(t)$ and $\bar{H}(t)\equiv y_{{\vec{r}'}}(t)\hat O^{-1}  \sigma_z \otimes B(t)\hat O=f_{\hat o}^z y_{{\vec{r}'}}(t)\sigma_z\otimes B(t)$. From \cite{pazsilva2017} we know the cumulants can be calculated as
\begin{align*}
&\quad\mathcal{C}^{(1)}_{\hat o,\vec{r},{\vec{r}'}}(T)\\
&=\int_{-T}^T dt H_{\hat o,\vec{r},{\vec{r}'}}(t)= \int_{-T}^0dt \tilde{H}(T+t)-\int_0^Tdt \bar{H}(T-t)\\
&=\!\int_0^T\!\! dt\braket{\tilde{H}(t)\!-\!\bar{H}(t)} =\!\int_0^T \!\! dt[y_{\vec{r}}(t)\!-\!y_{{\vec{r}'}}(t) f_{\hat o}^z]\braket{\sigma_z\!\otimes\! B(t)}\\
&=2\sigma_z\int_0^T dt Y^{-,\hat o}_{\vec{r},\vec{r}'}(t)\langle B(t)\rangle=0,
\end{align*}
\begin{widetext}
\begin{align*}
\mathcal{C}^{(2)}_{\hat o,\vec{r},{\vec{r}'}}(T)&=2\int_{-T}^T dt \int_{-T}^{t} dt' \braket{H_{\hat o,\vec{r},{\vec{r}'}}(t)H_{\hat o,\vec{r},{\vec{r}'}}(t')}-\mathcal{C}^{(1)}_{\hat o,\vec{r},{\vec{r}'}}(T)^2\\
&=2\Big(\int_{-T}^0 dt \int_{-T}^{t} dt'+\int_{0}^T dt \int_{-T}^{0} dt'+\int_{0}^T dt \int_{0}^{t} dt'\Big)\braket{H_{\hat o,\vec{r},{\vec{r}'}}(t)H_{\hat o,\vec{r},{\vec{r}'}}(t')}\\
&=2\int_0^T dt\int_0^{t} dt'\braket{\tilde{H}(t)\tilde{H}(t')+\bar{H}(t')\bar{H}(t)-\bar{H}(t)\tilde{H}(t')-\bar{H}(t')\tilde{H}(t)}\\
&=2I_S \int_0^T dt\int_0^{t} dt' y_{\vec{r}}(t)y_{\vec{r}}(t') \braket{B(t)B(t')}+f_{\hat o}^z y_{\vec{r}'}(t')f_{\hat o}^z y_{\vec{r}'}(t) \braket{B(t')B(t)}-f_{\hat o}^z y_{\vec{r}'}(t)y_{\vec{r}}(t') \braket{B(t)B(t')}\\
&\quad -f_{\hat o}^z y_{\vec{r}'}(t')y_{\vec{r}}(t) \braket{B(t')B(t)}\\
&=4I_S \int_0^T dt\int_0^{t} dt' Y^{-,\hat o}_{\vec{r},\vec{r}'}(t)[Y^{-,\hat o}_{\vec{r},\vec{r}'}(t')+Y^{+,\hat o}_{\vec{r},\vec{r}'}(t')]\braket{B(t)B(t')}+Y^{-,\hat o}_{\vec{r},\vec{r}'}(t)[Y^{-,\hat o}_{\vec{r},\vec{r}'}(t')-Y^{+,\hat o}_{\vec{r},\vec{r}'}(t')]\braket{B(t')B(t)}\\
&=4I_S \int_0^T dt\int_0^{t} dt' Y^{-,\hat o}_{\vec{r},\vec{r}'}(t)Y^{-,\hat o}_{\vec{r},\vec{r}'}(t')\braket{[B(t),B(t')]_+}+Y^{-,\hat o}_{\vec{r},\vec{r}'}(t)Y^{+,\hat o}_{\vec{r},\vec{r}'}(t')\braket{[B(t),B(t')]_-}\\
&= 2 I_S\int_0^T dt\int_0^T dt'  Y_{\vec{r},\vec{r}'}^{-,\hat o}(t) Y_{\vec{r},\vec{r}'}^{-,\hat o}(t') \langle{[B(t),B(t')]_+}\rangle  +4I_S \int_0^T dt\int_0^{t} dt' Y_{\vec{r},\vec{r}'}^{-,\hat o}(t) Y_{\vec{r},\vec{r}'}^{+,\hat o}(t') \langle{[B(t),B(t')]_-}\rangle.
\end{align*}
\end{widetext}

\section{Procedures to efficiently exploit Eq. (\ref{eqa6})} 
\label{sec:prac}

When exploiting Eq.~\eqref{eqa6} to estimate ${\tiny{\begin{pmatrix}
\cdot & \cdot & 0\\
\cdot & 0 & 0\\
0 & 0 & 1
\end{pmatrix}}}^+$, a necessary condition for the estimation result to be reliable is that $4\mathcal{B}^2-\mathcal{A}^2$ should not be too close to zero (otherwise the calculated $\sinh$ function value from Eq.~\eqref{eqa6} may deviate significantly from the true value). If one directly tests the value of $4\mathcal{B}^2-\mathcal{A}^2$ based on Eq.~\eqref{eqa6}, it is neither efficient nor accurate, because there are altogether six expectation values of observables under different system configurations to be measured. Here we introduce an approach to efficiently test the value of $4\mathcal{B}^2-\mathcal{A}^2$ under any effective switching function chosen/to test, denoted as $Y^{\pm,\text{chosen/test}}_{\vec{a}}(t)$. 

We start from the expression of $4\mathcal{B}^2-\mathcal{A}^2$ as
\begin{align}
&\quad 4\mathcal{B}^2-\mathcal{A}^2\notag\\
&=16\cos^2\Big(2{\scriptsize{\begin{pmatrix}
\cdot & \cdot & 0\\
\cdot & 0 & 1\\
0 & 0 & 0
\end{pmatrix}}^-}\Big)\Big[A_{(0,0,1)}^+\Big]^2\Big[A_{(1,0,0)}^+\Big]^2.\label{eqa5}
\end{align}

(i) We note that based on Eq.~\eqref{eqb4},  the value of $A_{(1,0,0)}^+$ can be obtained from an $N=1$ experiment of $E [\hat{Y}(t_1)]_{|+\rangle_y}$. After a pair of candidate $Y^{-,\text{test}}_{(1,0,0)}(t)$, possibly with a traditional switching function ($\pi$-pulse sequence) to be tested encoded inside the first interval, and $t_1^{\text{test}}$ are proposed based on certain reconstruction algorithm, e.g., the one in Sec. \ref{subsec:fouriertrans}, one can experimentally evaluate the value of $E [\hat{Y}(t_1^{\text{test}})]_{|+\rangle_y}$, with $Y^{-,\text{test}}_{(1,0,0)}(t)$ modulating the first interval $[0,t_1]$. If the obtained expectation value decoheres, $Y^{-,\text{test}}_{(1,0,0)}(t)$ and $t_1^{\text{test}}$ are not appropriate settings for the specific noise environment.

\begin{figure}
\centering
\includegraphics[scale=0.4]{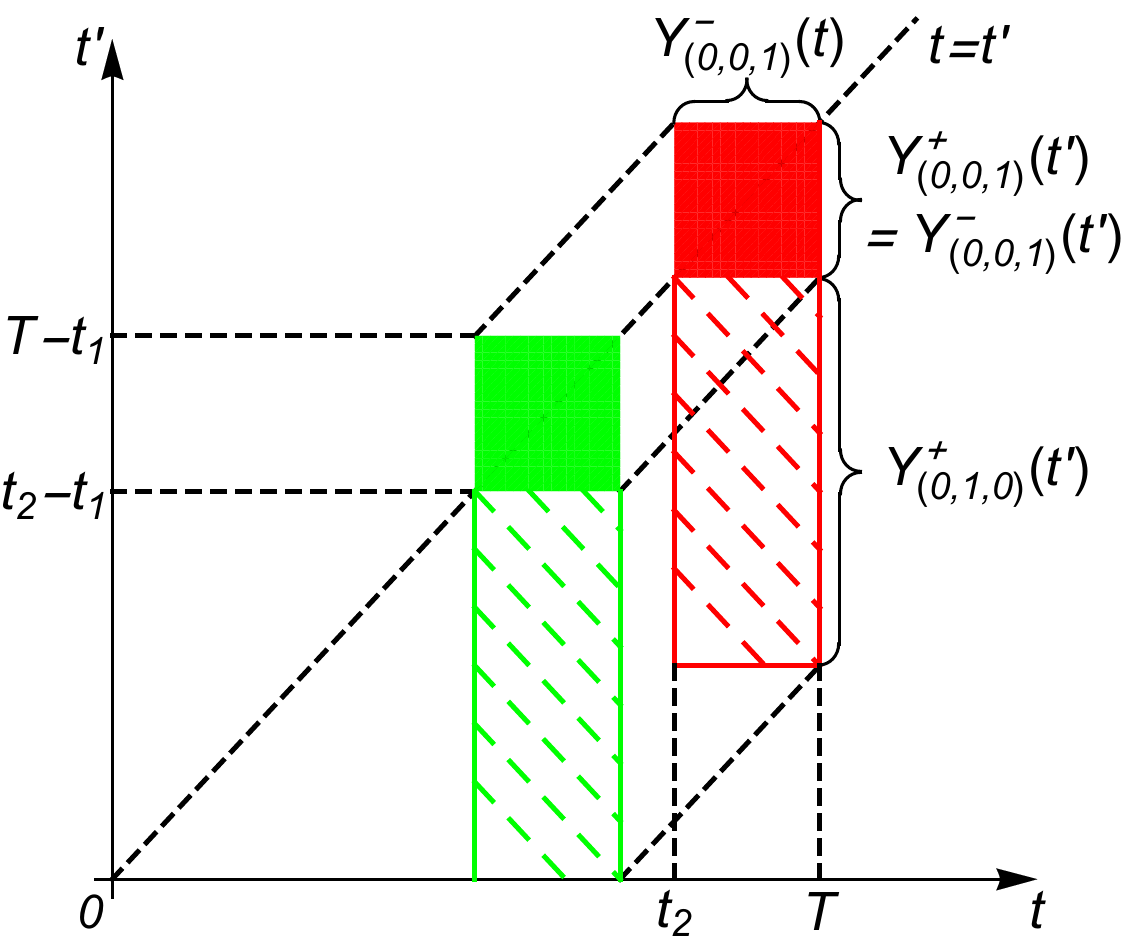}
\caption{Accessing the target noise block using stationarity. The quantity of Eq.~\eqref{eqx07} corresponds to the red solid and dashed blocks, which can be moved to the green blocks directly obtained via a measurement of properly configured Eq. \eqref{eqb14}.}
\label{fsmall}
\end{figure}

(ii) The remaining undetermined quantity in Eq.~\eqref{eqa5} is the square of 
\begin{align}\label{eqx07}
\cos(2{\scriptsize{\begin{pmatrix}
\cdot & \cdot & 0\\
\cdot & 0 & 1\\
0 & 0 & 0
\end{pmatrix}}^-})A_{(0,0,1)}^+,
\end{align}
which should not be close to zero. One way to control the magnitude of the $\cos(\cdot)$ component is to choose $Y_{(0,1,0)}^{+,\text{test}}(t)$ to be a DD sequence (the second interval $[t_1,t_2]$ is absent in the RHS of Eq.~\eqref{eqa6}, so we have the freedom to determine its $\pi$-pulse sequence), such that $\cos(2{\scriptsize{\begin{pmatrix}
\cdot & \cdot & 0\\
\cdot & 0 & 1\\
0 & 0 & 0
\end{pmatrix}}^-})$ is not close to zero. As for the third interval, if it shares the same $\pi$-pulse sequence with the first interval, we then know by stationarity that $A_{(0,0,1)}^{+,\text{test}}=A_{(1,0,0)}^{+,\text{chosen}}$, which is already determined. Otherwise if $[t_2,T]$ employs a different $\pi$-pulse sequence or interval length, then we notice that the two terms in Eq. \eqref{eqx07} are the red solid and dashed blocks in Fig. \ref{fsmall}, which can be moved to the green blocks using stationarity. Hence, by performing an $N=2$ experiment of Eq. \eqref{eqb14} matching the green configurations in Fig. \ref{fsmall} and employing $Y_{(0,1,0)}^{+,\text{test}}(t)$ and $Y_{(0,0,1)}^{-,\text{test}}(t)$ sequences in the first and second intervals respectively, one can immediately obtain the value of Eq. \eqref{eqx07} and confirm the applicability of $Y_{(0,1,0)}^{+,\text{test}}(t)$, $t_2^{\text{test}}$ and $Y_{(0,0,1)}^{-,\text{test}}(t)$.

Now after measuring only two expectation values of observables (possibly each repeated for several rounds to determine a proper setting), we can determine the magnitude of Eq. \eqref{eqa5} to ensure it is large enough.

\section{How to determine the sampling setting}\label{app:para}
Assuming that the different $t_2$ values (which we call \textit{time traces}, similar to the usage in \cite{zhang2014}) are equidistant such that $t_2=kT_s|_{k=0,1,2,\cdots, K}$. Under the assumption of infinite measurement shots for each observable, given certain $F(\omega,t_1)$, one is sampling the function $\mathcal{I}(t_2)^\pm=\mathcal{I}(t_1,t_2,t_1+t_2)^\pm$ with period $T_s$, obtaining the values of $\mathcal{I}(kT_s)^\pm$, to reconstruct the Fourier transform of $\mathcal{I}(t_2)^\pm$ in the frequency region. Utilizing the symmetry/antisymmetry of $\mathcal{I}(t_2)^\pm$, one can extend $k$ to $\{-K,\cdots,0,\cdots,K\}$. Here we analyze how the values of $K$ and $T_s$ affect the reconstruction accuracy, by reviewing the procedures of proving the Nyquist–Shannon sampling theorem in, e.g., \cite{nassar}.

One can view the sampling results as the product of the target function $\mathcal{I}(t_2)^\pm$ with an impulse-train function $\delta_{T_s}^K(t)\equiv \sum_{k=-K}^K \delta(t-kT_s)$. Its Fourier series expansion is $\delta_{T_s}^\infty(t)=\sum_{n=-\infty}^{\infty}e^{-{\rm i}n{\omega_2}t}/T_s$, where $\omega_2\equiv 2\pi/T_s$. By the convolution ($\ast$) theorem, the Fourier transform of the practical data is 
\begin{align}
&\quad \mathcal{F}[\mathcal{I}(t_2)^\pm\cdot \delta_{T_s}^K(t)]\label{eqf16}\\
&=\mathcal{F}[\mathcal{I}(t_2)^\pm]\ast\mathcal{F}[\delta_{T_s}^K(t)]/2\pi\notag\\
&=\frac{1}{2\pi}\int_{-\infty}^\infty d\tau |F(\omega-\tau)|^2S^\pm(\omega-\tau)\sum_{k=-K}^K e^{-{\rm i}\tau kT_s} . \notag
\end{align}
When ideally there are infinite time traces ($K=\infty$), $\sum_{k=-K}^K e^{-{\rm i}\tau kT_s}=\omega_2 \delta_{\omega_2}^\infty(\tau)$. Hence, \eqref{eqf16} becomes
\begin{align}
&\quad \mathcal{F}[\mathcal{I}(t_2)^\pm\cdot \delta_{T_s}^\infty(t)]\notag\\
&=\frac{1}{2\pi}\int_{-\infty}^\infty d\tau |F(\omega-\tau)|^2S^\pm(\omega-\tau)\omega_2 \delta_{\omega_2}^\infty(\tau) \notag\\
&=\frac{1}{T_s}\sum_{k=-\infty}^\infty |F(\omega-k\omega_2)|^2S^\pm(\omega-k\omega_2),\label{eqf17}
\end{align}
which repeats shifting the original filtered spectrum $|F(\omega)|^2S^\pm (\omega)$ by a distance of multiple $\omega_2$, adds all the images together and divides them by $T_s$. If there is a high frequency cutoff $\omega_c$ for $|F(\omega)|^2S^\pm(\omega)$, then $\omega_2>2\omega_c$ guarantees that the shifted images do not overlap/alias and a perfect recover of the original spectrum is allowed in principle, which is the famous sampling theorem. Differently, here we are only interested in the MFS $[\omega_a,\omega_b]$ and allow frequency aliasing to happen outside MFS. Therefore, we only require $\omega_2>\omega_c+\omega_b$, i.e.,
\begin{equation}\label{eqAppC3}
    T_s<\frac{2\pi}{\omega_c+\omega_b}.
\end{equation}
The filter function is typically equipped with high frequency lobes, causing spectral leakage \cite{norris2018}. An ideal $\omega_c$ thus does not really exist, and in practice one can take $\omega_c$ to be the first positive root of $|F(\omega)|^2$ or a sufficiently large value. Moreover, there are potential tricks worth developing. For example, (i) special $T_s$ values can be chosen such that clean regions (where $|F(\omega)|^2$ is basically zero and without non-negligible lobes) of the filter function cover the MFS, or (ii) different groups of $T_s$ values can be employed such that the position of the lobes differ from groups to groups and the reconstruction results in the unaffected subregions of MFS can be pieced together. We do not delve into these technical details in this paper and only point out the potentials.

Assuming that $T_s$ is already properly chosen such that the frequency aliasing problem is solved, we illustrate the effect of a finite $K$. If $K$ is not large enough, $\mathcal{F}[\delta_{T_s}^K(t)]$ will not be close enough to $\omega_2 \delta_{\omega_2}^\infty(\tau)$, distorting the ideal result of \eqref{eqf17}. Hence, by depicting the plot of the Dirichlet kernel $\sum_{k=-K}^K e^{-{\rm i}\tau kT_s}=\sin[\frac{\tau T_s}{2}(2K+1)]/\sin(\frac{\tau T_s}{2})$ versus the ideal $\omega_2 \delta_{\omega_2}^\infty(\tau)$ in the regime $0\leq \tau\leq \omega_a$, one can roughly estimate whether certain $K$ value is large enough. A natural intuitive requirement is that the first positive root of $\sin[\frac{\tau T_s}{2}(2K+1)]/\sin(\frac{\tau T_s}{2})=0$ should be significantly small compared with $\omega_a$, which reads 
\begin{equation}\label{eqf4}
\frac{2\pi}{T_s(2K+1)}\ll \omega_a.    
\end{equation}
Hence, there is a trade-off between $T_s$ and $K$: for a given filter function $F(\omega,t_1)$, the smaller is the sampling period $T_s$, the larger is the needed sampling time trace number $K$. Although increasing the sampling frequency mitigates the frequency aliasing problem, the price is that more resources (time traces) are needed.

A special scenario breaking Eq. \eqref{eqf4} is when $\omega_a=0$, i.e., a free evolution filter function is employed whose MFS covers $\omega=0$, especially for 1/f spectra. Related to this, an important observation is that the larger is $K$, the better does the reconstructed spectrum capture the peaks or steep slopes on the spectral curve. Imagine that $T_s$ is small enough such that the filtered spectrum is mainly supported in $[-\omega_2/2,\omega_2/2]$ such that the aliasing error can be neglected. Assuming that the filtered spectrum has a finite-term Fourier series expansion in this region as
\begin{equation}
    |F(\omega)|^2S^\pm(\omega)|_{\omega\in[-\omega_2/2,\omega_2/2]}=\sum_{n=-N}^N c_n e^{{\text i}n\omega T_s}.
\end{equation}
Then \eqref{eqf16} becomes
\begin{equation}\label{eqf27}
\begin{aligned}
&\quad \mathcal{F}[\mathcal{I}(t_2)^\pm\cdot \delta_{T_s}^K(t)]|_{\omega\in[-\omega_2/2,\omega_2/2]}\\
&=\frac{1}{2\pi}\int_{-\infty}^\infty d\tau \sum_{n=-N}^N c_n e^{{\text i}n(\omega-\tau)T_s}\sum_{k=-K}^K e^{-{\rm i}\tau kT_s}\\
&=\frac{1}{2\pi} \sum_{n=-N}^N c_n e^{{\text i}n\omega T_s}\sum_{k=-K}^K \int_{-\infty}^\infty d\tau e^{{\text i}\tau (-k-n)T_s}\\
&=\sum_{n=-N}^N c_n e^{{\text i}n\omega T_s}\sum_{k=-K}^K \delta[(-k-n)]/T_s.
\end{aligned}
\end{equation}
When $K\geq N$, Eq.
\eqref{eqf27}$\propto |F(\omega)|^2S^\pm(\omega)|_{\omega\in[-\omega_2/2,\omega_2/2]}$ and $S^\pm (\omega)|_{\omega\in[-\omega_2/2,\omega_2/2]}$ can be correctly reconstructed in theory. Otherwise when $K<N$, \eqref{eqf27} contains no information about the high-order Fourier terms $\sum_{K+1\leq |n|\leq N} c_n e^{{\text i}n\omega T_s}$ constituting the peaks on the spectrum curve. Hence, when the time trace sequence is not long enough, the peaks or steep slopes in the spectrum curve usually might not be fully captured.

Denote $\Delta S_A^\pm(\omega)\equiv \sum_{k\neq 0} |F(\omega-k\omega_2)|^2S^\pm(\omega-k\omega_2)/|F(\omega)|^2$ the aliasing error, in the sense of 
\begin{equation}\label{eqf23}
S^\pm(\omega)=\frac{T_s}{|F(\omega)|^2}\mathcal{F}[ \mathcal{I}(t)^\pm\delta_{T_s}^\infty(t)]-\Delta S_A^\pm(\omega)
\end{equation}
from \eqref{eqf17}. Assuming that the aliasing error and finite-time-trace error are neglected by the spectrum reconstruction algorithm (DTFT here) while they still exist in the actual physics process, we now analyze the net error, including the sampling error $\hat{\mathcal{I}}(t_2)^\pm-\mathcal{I}(t_2)^\pm$, e.g., from finite sampling (i.e., finite statistics when measuring each observable). The spectrum estimation error can be bounded as
\begin{equation}\label{eqf21}
\begin{aligned}
&\quad |S^\pm(\omega)- \hat S^\pm(\omega)|\\
&=\Big|\frac{T_s}{|F(\omega)|^2}\mathcal{F}[ \mathcal{I}(t)^\pm\delta_{T_s}^\infty(t)]-\Delta S_A^\pm(\omega)\\
&\quad -\frac{T_s}{|F(\omega)|^2}\mathcal{F}[\hat{\mathcal{I}}(t)^\pm\delta_{T_s}^K(t)]\Big|\\
&\leq \Big|\mathcal{F}[ \mathcal{I}(t)^\pm\delta_{T_s}^\infty(t)]-\mathcal{F}[ \mathcal{I}(t)^\pm\delta_{T_s}^K(t)]\Big|\frac{T_s}{|F(\omega)|^2}\\
&\quad +\Big|\mathcal{F}[ \mathcal{I}(t)^\pm\delta_{T_s}^K(t)]-\mathcal{F}[\hat{\mathcal{I}}(t)^\pm\delta_{T_s}^K(t)]\Big|\frac{T_s}{|F(\omega)|^2}\\
&\quad +|\Delta S_A^\pm(\omega)|\\
&= \Big|\mathcal{F}[\mathcal{I}(t)^\pm(\delta_{T_s}^\infty(t)-\delta_{T_s}^K(t))]\Big|\frac{T_s}{|F(\omega)|^2} +|\Delta S_A^\pm(\omega)| \\
&\quad +\Big|\int dt [\mathcal{I}(t)^\pm-\hat{\mathcal{I}}(t)^\pm]\delta_{T_s}^K(t)e^{-{\rm i}\omega t}\Big|\frac{T_s}{|F(\omega)|^2},
\end{aligned}
\end{equation}
where on the RHS of the last equality the first term is the spectrum estimation error from finite time traces, the second term is the aliasing error and the third term can be bounded as
\begin{equation}\label{eqf22}
    \begin{aligned}
    &\quad\Big|\int dt [\mathcal{I}(t)^\pm-\hat{\mathcal{I}}(t)^\pm]\delta_{T_s}^K(t)e^{-{\rm i}\omega t}\Big|\frac{T_s}{|F(\omega)|^2}\\
&\leq \int  dt \Big|[ \mathcal{I}(t)^\pm-\hat{\mathcal{I}}(t)^\pm]\delta_{T_s}^K(t)e^{-{\rm i}\omega t}\Big|\frac{T_s}{|F(\omega)|^2}\\
&=\sum_{k=-K}^K | \mathcal{I}(kT_s)^\pm-\hat{\mathcal{I}}(kT_s)^\pm| \frac{T_s}{|F(\omega)|^2}\\
&=\frac{T_s}{|F(\omega)|^2}\sum_{k=0}^K (2-\delta_{k0})| \mathcal{I}(kT_s)^\pm-\hat{\mathcal{I}}(kT_s)^\pm|,
\end{aligned}
\end{equation}
representing the estimation error from the inaccuracy of each recorded time trace. Eqs. \eqref{eqf21} and \eqref{eqf22} give an error upper bound separating different error sources. Moreover, \eqref{eqf22} can be linked with finite sampling through error propagation formula, bounding the error contribution from there and assisting in determining proper measurement shot numbers. It is also possible to mitigate the aliasing error using \eqref{eqf23} based on the existing estimation results. We do not elaborate on these potential developments in this paper.

\section{Proof of Theorem \ref{theorem2}}\label{app:theorem}

\begin{proof}
To prove Eq. \eqref{eq41}, firstly we will show that it suffices to assume $\rho_B$ is pure. Define 
\begin{align*}
&\quad G^\pm(\rho_B)=G^\pm(\rho_B,\omega)\\
&\equiv \int_{-\infty}^{\infty}dt\  e^{-{\rm i}\omega t}\Tr_B\Big(\langle[B(t),B(0)]_+\rangle_c \rho_B\Big)\\
&\quad \pm \int_{-\infty}^{\infty}dt\  e^{-{\rm i}\omega t}\Tr_B\Big(\langle[B(t),B(0)]_-\rangle_c \rho_B\Big),
\end{align*}
where $\omega\in\mathbb{R}$. Clearly $G^\pm(\rho_B)$ are two real linear functionals defined over the space of all the quantum states on $\mathcal{H}$. Since mixed states are convex combinations of pure states, to prove Eq. \eqref{eq41}, it suffices to prove that $G^\pm(\rho_B)\geq 0$ hold for any pure $\rho_B$, given arbitrary $\omega\in\mathbb{R}$. We thus assume $\rho_B=|r\rangle\langle r|$.

Next, take a resolution of identity $I=\int_P dp|p\rangle\langle p|$ such that $|r\rangle\notin\{|p\rangle\}_p$ (but we allow $|r\rangle\in \text{span}\{|p\rangle\}_p$). For any $|a\rangle,|b\rangle\in\mathcal{H}$, denote $B_{ab}(t)= x_{ab}(t)-{{{\rm i}}}y_{ab}(t)$ where $x_{ab}(t)=x_{ba}(t)$ and $y_{ab}(t)=-y_{ba}(t)$ are classical wide-sense stationary stochastic processes. We then know
\begin{align}
&G^+(|r\rangle\langle r|,\omega)\notag\\
=&\int_{-\infty}^{\infty}dt\  e^{-{\rm i}\omega t}\Big(\Tr_B\{\langle[B(t),B(0)]_+\rangle_c \rho_B\}\notag\\
&\quad +\Tr_B\{\langle[B(t),B(0)]_-\rangle_c \rho_B\}\Big)\notag \\
=&2\int_{-\infty}^{\infty}dt\  e^{-{\rm i}\omega t}Tr_B\Big(\langle B(t)B(0)\rangle_c \rho_B\Big)\notag\\
=&2\int_{-\infty}^{\infty}dt\  e^{-{\rm i}\omega t}\int_P dp\langle\langle r|B(t)|p\rangle\langle p|B(0)|r\rangle\rangle_c\notag \\
=&2\!\int_{-\infty}^{\infty}dt\  e^{-{\rm i}\omega t}\!\int_P dp\langle [x_{r p}(t)\!-\!{{\rm i}}y_{r p}(t)][ x_{pr }(0)\!-\!{{\rm i}}y_{pr }(0)]\rangle_c\notag \\
=&2\int_{-\infty}^{\infty}dt\  e^{-{\rm i}\omega t}\int_P dp\langle x_{r p}(t)x_{r p}(0)+y_{r p}(t)y_{r p}(0)\notag\\
&\quad +{{\rm i}}x_{r p}(t)y_{r p}(0)-{{\rm i}}x_{r p}(0)y_{r p}(t)\rangle_c.
\label{eq47}
\end{align}

Since one of Eq. \eqref{eq39} holds, using Fubini–Tonelli theorem, we can swap the double integrals in Eq. \eqref{eq47} to reach
\begin{align*}
&\quad G^+(|r\rangle\langle r|,\omega)\\
&=2\!\int_P  dp\Big( S_{rp}^{x,+}(\omega)+S_{rp}^{y,+}(\omega) +{{\rm i}}S_{rp}^{xy,\times}(\omega)-{{\rm i}}S_{rp}^{yx,\times}(\omega)\Big).
\end{align*}
Denote the classical spectrum of $x_{ab}(t)$ and $y_{ab}(t)$ to be $S_{ab}^{x,+}(\omega)$ and $S_{ab}^{y,+}(\omega)$, respectively. Let the cross spectrum of $\langle x_{ab}(t)y_{ab}(t')\rangle_c$ be $S_{ab}^{xy,\times}(\omega)$, which can be complex. A key ingredient is the {\textit{cross-spectrum inequality}} in classical signal processing (see e.g., Eq. (5.89b) of \cite{bendat}) as
\begin{equation}\label{eqb46}
|S_{ab}^{xy,\times}(\omega)|^2\leq S_{ab}^{x,+}(\omega)S_{ab}^{y,+}(\omega).
\end{equation}
We thus know
\begin{align*}
&\quad G^+(|r\rangle\langle r|,\omega)\\
&\geq 2\!\int_P  \!dp\Big(\! S_{rp}^{x,+}(\omega)+S_{rp}^{y,+}(\omega) -|S_{rp}^{xy,\times}(\omega)|-|S_{rp}^{yx,\times}(\omega)|\Big)\\
&\geq 2\int_P  dp\Big( S_{rp}^{x,+}(\omega)+S_{rp}^{y,+}(\omega) - 2\sqrt{S_{rp}^{x,+}(\omega)S_{rp}^{y,+}(\omega)}\Big)\\
&\geq 0.
\end{align*}
Similarly, one can prove $G^-(|r\rangle\langle r|,\omega)\geq 0$, which completes the proof of Eq. \eqref{eq41}.

For Eq. \eqref{eq40}, based on Eq.  \eqref{eq41} we know
\begin{align*}
&\quad \Big|\langle[B(t),B(t')]_-\rangle\Big|=\frac{1}{2\pi}\Big|\int_{-\infty}^\infty S^-(\omega)e^{{\rm i}\omega (t-t')}d\omega\Big|\\
&\leq \frac{1}{2\pi}\int_{-\infty}^\infty \Big|S^-(\omega)\Big|d\omega \leq \frac{1}{2\pi}\int_{-\infty}^\infty S^+(\omega)d\omega\\
&=\langle[B(0),B(0)]_+\rangle.
\end{align*}

Denote $F_0(\omega,T)\equiv\int_0^T e^{{\rm i}\omega t}dt$ the filter function generated by free evolution. For Eq. \eqref{eq42}, based on Eq. \eqref{eq41} we know
\begin{align*}
&\quad  2\Big|\int_{T_1}^{T_1+T_2}dt\int_0^{T_1}dt' \langle[B(t),B(t')]_-\rangle\Big|\\
&=2\Big|\int_0^{T_2}d\bar{t}\int_{0}^{T_1}dt' \langle [B(T_1+\bar{t}-t'),B(0)]_-\rangle\Big|\\
&=\frac{2}{2\pi}\Big|\int_0^{T_2}d\bar{t}\int_{0}^{T_1}dt' \int_{-\infty}^{\infty}d\omega S^-(\omega)e^{\text{i}\omega(T_1+\bar{t}-t')}\Big|\\
&=\frac{1}{\pi}\Big|\int_{-\infty}^\infty d\omega F_0(\omega,T_1)^*F_0(\omega,T_2)e^{{\rm i}\omega T_1}S^-(\omega)\Big|\\
&\leq \frac{1}{\pi}\int_{-\infty}^\infty d\omega\ |F_0(\omega,T_2)||F_0(\omega,T_1)||S^{-}(\omega)|\\
&\leq \frac{1}{2\pi}\int_{-\infty}^\infty d\omega\Big[|F_0(\omega,T_1)|^2+|F_0(\omega,T_2)|^2\Big]S^+(\omega)\\
&=\frac{1}{2\pi}\int_{-\infty}^\infty S^+(\omega)d\omega\Big(\int_0^{T_1}dt\int_0^{T_1}dt'e^{{\rm i}\omega (t-t')}\\
&\quad +\int_0^{T_2}dt\int_0^{T_2}dt'e^{{\rm i}\omega (t-t')}\Big)\\
&=\Big(\int_0^{T_1}dt\int_0^{T_1}dt'+\int_0^{T_2}dt\int_0^{T_2}dt'\Big)\langle[B(t),B(t')]_+\rangle.
\end{align*}

As for the scenario where $\hat S^\pm(\omega)$ are given as illustrated, we take $\hat{\mathcal{H}}$ to be the common position-momentum space (in $\mathbb{R}^1$ instead of $\mathbb{R}^3$) with a resolution of identity $I=\int_{\mathbb{R}} dp |p\rangle\langle p|$. As before, from $\text{span}\{|p\rangle\}_p$ we take $|r\rangle$ such that $|r\rangle\notin\{|p\rangle\}_p$. Let $\{a_p\}$, $\{b_p\}$, $\{c_p\}$ and $\{d_p\}$ be four independent white noise processes indexed by $p\in\mathbb{R}$, all with zero means and unit variances. We then define
\begin{align*}
\hat x_{rp}(t)\!&\equiv\!\frac{1}{2\sqrt{2\pi}} {\rm{sgn}}(\hat S^-\!(p))\sqrt{{|\hat S^-\!(p)|}}\Big[\!a_p\cos(p t)\!+\!b_p\sin(p t)\!\Big]\\
&\quad +\!\frac{1}{2\sqrt{\pi}} \sqrt{{\hat S^+(p)\!-\!|\hat S^-(p)|}}\Big[c_p\cos(pt)\!+\!d_p\sin(pt)\Big]\!,\\
\hat y_{rp}(t)\!&\equiv\!\frac{1}{2\sqrt{2\pi}}\sqrt{{|\hat S^-(p)|}}\Big[b_p\cos(p t)-a_p\sin(p t)\Big],
\end{align*}
which have correlation functions
\begin{align}
&\quad\langle \hat x_{rp}(t)\hat x_{rp}(t')\rangle_c\notag \\ &=\frac{1}{4\pi}\Big[\hat S^+(p)\!-\!|\hat S^-(p)|\Big]\Big[\!\cos(p t)\cos(p t')\!+\!\sin(p t)\sin(p t')\!\Big]\notag\\
&\quad + \frac{1}{8\pi}|\hat S^-(p)|\Big[\cos(p t)\cos(p t')+\sin(p t)\sin(p t')\Big] \notag \\
&=\frac{1}{8\pi}\Big[2\hat S^+(p)-|\hat S^-(p)|\Big]\cos(p t-p t'),
\label{eqB14}
\end{align}
and
\begin{align}\label{eqB15}
\langle \hat y_{rp}(t)\hat y_{rp}(t')\rangle_c =\frac{1}{8\pi}|\hat S^-(p)|\cos(p t-p t').
\end{align}
Therefore, both $\hat x_{rp}(t)$ and $\hat y_{rp}(t)$ are zero-mean wide-sense stationary processes. We take the bath operator such that $\hat B_{rp}(t)=\hat x_{rp}(t)-{\rm i} \hat y_{rp}(t)$. Using Eq. \eqref{eq47}, we know
\begin{align*}
& \quad \langle [\hat B(t),\hat B(0)]_-\rangle\!=\!\left\langle Tr_B\Big([\hat B(t)\hat B(0)-\hat B(0)\hat B(t)]\rho_B\Big)\right\rangle_c \\
&=2\int_{\mathbb{R}} dp\langle {{\rm i}}\hat x_{r p}(t)\hat y_{r p}(0)-{{\rm i}}\hat x_{r p}(0)\hat y_{r p}(t)\rangle_c\\
&=\frac{2{\rm i}}{8\pi}\int_{\mathbb{R}} dp\ \hat S^-(p)\Big[\sin(pt) +\sin(pt)\Big]\\
&=\frac{1}{2\pi}\int_{\mathbb{R}} dp\ \hat S^-(p)e^{{\rm i} pt}=\mathcal{F}^{-1}\Big[\hat S^-(p)\Big],
\end{align*}
and together with Eqs. \eqref{eqB14} and \eqref{eqB15} we know
\begin{align*}
& \quad \langle [\hat B(t),\hat B(0)]_+\rangle\!=\!\left\langle Tr_B\Big([\hat B(t)\hat B(0)+\hat B(0)\hat B(t)]\rho_B\Big)\right\rangle_c\\
&=2\int_{\mathbb{R}} dp\langle \hat x_{r p}(t)\hat x_{r p}(0)+\hat y_{r p}(t)\hat y_{r p}(0)\rangle_c\\
&=\frac{2}{8\pi}\int_{\mathbb{R}} dp\ 2\hat S^+(p)\cos(pt) =\frac{1}{2\pi}\int_{\mathbb{R}} dp\ \hat S^+(p)e^{{\rm i} pt}\\
&=\mathcal{F}^{-1}\Big[\hat S^+(p)\Big],
\end{align*}
which completes the proof.
\end{proof}

\bibliographystyle{alpha}

\end{document}